\documentstyle[prd,aps,floats,tighten,epsfig,graphicx,color]{revtex}

\newcommand{\beq}{\begin{equation}}
\newcommand{\eeq}{\end{equation}}
\newcommand{\bea}{\begin{eqnarray}}
\newcommand{\ena}{\end{eqnarray}}
\newcommand{\etal}{{\it et al.}}

\newcommand{\lsim}{\mathrel{\mathop{\kern 0pt \rlap
{\raise.2ex\hbox{$<$}}}
\lower.9ex\hbox{\kern-.190em $\sim$}}}
\newcommand{\gsim}{\mathrel{\mathop{\kern 0pt \rlap
{\raise.2ex\hbox{$>$}}}
\lower.9ex\hbox{\kern-.190em $\sim$}}}

\newcommand{\astroph}[1]{{\tt astro-ph/#1}}

\renewcommand{\apj}[3]{Astrophys.\ J.\ {\bf #1}, #3 (#2)}

\newcommand{\href}[2]{#1}

\newcommand{\msol}{\mbox{$M_{\odot}$}}

\newcommand{\omegaM}{\mbox{$\Omega_{\rm M}$}}
\newcommand{\omegaB}{\mbox{$\Omega_{\rm B}$}}
\newcommand{\omegaL}{\mbox{$\Omega_{\Lambda}$}}

\definecolor{cyan}{cmyk}{1.,0.,0.,0.5}
\definecolor{magenta}{cmyk}{0.,1.,0.,0.5}
\definecolor{verdatre}{cmyk}{0.5,0.,0.5,0.5}
\definecolor{yellow}{cmyk}{0.,0.,0.2,0.0}
\definecolor{rouge}{cmyk}{0.,0.4,0.6,0.0}
\definecolor{orange}{cmyk}{0.,0.5,0.5,0.}
\definecolor{violet}{rgb}{0.5,0.,0.5}

\begin{document}

\draft
\title{Quintessential Haloes around Galaxies}
\vskip 1.cm
\author{
Alexandre Arbey$^{\rm a,b}$
\footnote{E--mail: arbey@lapp.in2p3.fr, lesgourg@lapp.in2p3.fr,
salati@lapp.in2p3.fr},
Julien Lesgourgues$^{\rm a}$ and Pierre Salati$^{\rm a,b}$
}
\vskip 0.5cm
\address{
\begin{flushleft}
a) Laboratoire de Physique Th\'eorique LAPTH, B.P.~110, F-74941
Annecy-le-Vieux Cedex, France.\\
b) Universit\'e de Savoie, B.P.~1104, F-73011 Chamb\'ery Cedex,
France.
\end{flushleft}
}
\maketitle
\vskip 0.5cm
\centerline{11 September 2001}

\vskip 0.5cm
\begin{abstract}
The nature of the dark matter that binds galaxies remains
an open question. The favored candidate has been so far the
neutralino. This massive species with evanescent interactions
is now in difficulty. It would actually collapse in dense clumps
and would therefore play havoc with the matter it is supposed
to shepherd.
We focus here on a massive and non--interacting complex scalar
field as an alternate option to the astronomical missing mass. We
investigate the classical solutions that describe the Bose condensate
of such a field in gravitational interaction with matter. This simplistic
model accounts quite well for the dark matter inside low--luminosity
spirals whereas the agreement lessens for the brightest objects where
baryons dominate. A scalar mass $m \sim 0.4$ to $1.6 \times 10^{-23}$ eV is
derived when both high and low--luminosity spirals are fitted at the
same time. Comparison with astronomical observations is made
quantitative through a chi--squared analysis. We conclude that
scalar fields offer a promising direction worth being explored.
\end{abstract}
%
\vskip 1.cm

\section{Introduction.}
\label{sec:introduction}

The observations of the Cosmic Microwave Background
anisotropies \cite{boomerang} point towards a flat universe.
The determination of the relation between the distance of luminosity
and the redshift of supernovae SNeIa \cite{supernovae_omegaL}
strongly favors the existence of a cosmological constant which
contributes a fraction $\omegaL \sim 0.7$ to the closure density. The
pressure--to--density ratio $w$ of that fluid is negative with a value of
$w = - 1$ in the case of an exact cosmological constant. Alternatively,
this component could be in the form of dark energy
-- the so--called quintessence -- whose simplest incarnation is a neutral
scalar field $\Phi$ with the Lagrangian density
\beq
{\cal L} \; = \; \frac{1}{2} \, g^{\, \mu \nu} \,
\partial_{\mu}  \Phi \, \partial_{\nu}  \Phi
\, - \, V \left( \Phi \right) \;\; .
\label{scalar_neutral}
\eeq
Should the metric be flat and the field homogeneous, the energy
density may be expressed as
\beq
\rho \equiv T^{0}_{\; 0} \; = \; {\displaystyle \frac{\dot{\Phi}^{2}}{2}}
\, + \, V \left( \Phi \right) \;\; ,
\label{mass_density}
\eeq
whereas the pressure obtains from $T_{i j} \equiv - g_{\, i j} \, P$
so that
\beq
P \; = \; {\displaystyle \frac{\dot{\Phi}^{2}}{2}} \, - \, V \left( \Phi \right)
\;\; .
\label{pressure}
\eeq
If the kinetic term is negligible with respect to the contribution of the
potential, a pure cosmological constant -- $\omega = - 1$ -- is recovered.
Cosmological scenarios with quintessence in the form of a scalar field
have been investigated \cite{quintessence} with various potentials and
their relevance to structure formation has been discussed.

\vskip 0.1cm
On the other hand, matter contributes a fraction $\omegaM \sim 0.3$ to
the energy balance of the universe. The nature of that component is still
unresolved insofar as baryons amount only to \cite{nucleosynthese}
\beq
\omegaB \, h^{2} \; = \; 0.02 \pm 0.002 \;\; .
\eeq
According to the common wisdom, non--baryonic dark matter would
be made of neutralinos -- a massive species with weak interactions
that naturally arises in the framework of supersymmetric theories.
This approach has given rise to some excitement in the community
and many experimental projects have been developed to hunt for these
evading particles. The general enthusiasm has been recently refreshed
when numerical simulations have shown that cold dark matter would
cluster in very dense and numerous clumps \cite{moore} 
(see however \cite{Lin}). The halo of
the Milky Way should contain $\sim$ half a thousand satellites with
mass in excess of $10^{8}$ $\msol$ while a dozen only of dwarf--spheroidals
are seen. The clumps would also heat and eventually shred the galactic ridge.
More generally, this process would lead to the destruction of the disks of spirals.
A neutralino cusp would form at the centers of the latter. This is not
supported by the rotation curves of low surface brightness galaxies that
indicate on the contrary the presence of a core with constant density.
Finally, two--body interactions with halo neutralinos and its associated
dynamical friction would rapidly disrupt the otherwise observed spinning
bar at the center of the Milky Way.

\vskip 0.1cm
Neutralinos may be in jeopardy. New candidates are under scrutiny such as
particles with self interactions \cite{spergel}. An interesting possibility is based
on configurations of the above--mentioned scalar field $\Phi$ for which
the pressure $P$ vanishes.
An academic example is provided by the exponential potential
\beq
V \left( \Phi \right) \; = \; \frac{1}{2} \, \rho^{0}_{\Phi} \,
\exp \left\{ - \beta \left( \Phi_{0} - \Phi \right) \right\} \;\; ,
\eeq
where the parameter $\beta$ is
\beq
\beta \; = \;
{\sqrt{\displaystyle \frac{24 \, \pi \, G}{\Omega_{\Phi}}}} \;\; .
\eeq
In a flat and matter--dominated universe, such a field $\Phi$ would
behave just exactly as cold dark matter and would contribute a fraction
$\Omega_{\Phi} = \rho^{0}_{\Phi} / \rho^{0}_{C}$ to the closure density.
More generally, the kinetic energy ${\dot{\Phi}^{2}}/{2}$ should cancel
the potential $V \left( \Phi \right)$ in order for the pressure to vanish
and for the fluid to mimic the effect of non--relativistic matter. As
will be discussed in the next section, this is actually the case when
$\Phi$ behaves like an axion and oscillates coherently on a time scale
much shorter than the typical durations at stake. Alternatively, the
field $\Phi$ could have additional degrees of freedom and rotate in
the corresponding internal space.
The idea that the excess of gravity inside galaxies may be due to a
classical configuration of some scalar field $\Phi$ has already drawn
some attention. The discussion has nevertheless remained at an
introductory level. A condensate of massive bosons with repulsive
interparticle potential has been postulated \cite{goodman} to suppress
the formation of structure on subgalactic scales. The polytropic index
of this bosonic halo varies from $n = 1$ at low density up to $n = 3$
at high density. The stability and annihilation of such a system has
been mentioned in \cite{riotto_tkachev} and a limit on the quartic
coupling constant of $\Phi$ has been derived.
The relevance of scalar fields to the structure of galactic haloes and
their associated dark matter cannot be seriously addressed without
comparing the theoretical rotation curves to the observations. An
exponential potential -- with negative overall sign -- is shown
\cite{mexico} to lead to flat rotation curves. A massless and
non--interacting complex scalar field is thoroughly considered in
\cite{schunck}. The self--gravitating structure of the field is calculated.

\vskip 0.1cm
We nevertheless feel that these analysis may be improved in
several ways. To commence, a negative potential does not seem
quite appealing. It actually leads to unphysical situations where
the scalar field rolls down the hill indefinitely and converts
the infinite amount of energy stored into $V < 0$ into kinetic
energy. Then, as mentioned in \cite{nucamendi}, the rotation
curves are assumed to be flat up to an infinite distance. In both
\cite{mexico} and \cite{schunck}, the Bose condensate extends
to infinity and the mass of the system diverges linearly in the
radius $r$. The Minkowski metric is no longer recovered at
large $r$. On the contrary, space exhibits a small deficit of solid
angle. We feel that such a behavior is not realistic insofar as
the rotation curves of bright spirals are actually found to decrease
beyond their optical radius \cite{persic_salucci_stel}. Another
strange consequence is that Newton's gravitation does not
apply even when the fields are weak. Matching the metric with
the Robertson--Walker form may also be a problem. Finally,
the agreement between the predicted and the observed rotation
curves is only qualitative and in the case of \cite{mexico} is
based on just a few examples. The goodness of that agreement is
not assessed from a quantitative point of view.

\vskip 0.1cm
This motivated us to reinvestigate more thoroughly the subject.
In the next section, we discuss the general conditions under
which a scalar field may bind galaxies. We show that the same
field cannot easily account -- at the same time -- for the dark
matter at galactic scales and for the cosmological quintessence.
We explain the reasons which have lead us to consider the
model scrutinized in section~\ref{sec:the_model}. The
corresponding rotations curves are derived in section~\ref{sec:rc}
and are compared by means of a chi--squared analysis to the
universal curves unveiled by \cite{persic_salucci_stel}. The
results are discussed in section~\ref{sec:discussion} and prospects
for future investigations are finally suggested.

\newpage
\section{Can a scalar field bind galaxies ?}
\label{sec:generality}

In the weak field approximation of general relativity -- or
quasi--Newtonian limit -- deviations from the Minkowski metric
$\eta_{\mu \nu} = {\rm diag}(1,-1,-1,-1)$ are accounted for by the
perturbation $h_{\mu \nu}$. In the harmonic coordinate gauge
where it satisfies the condition
\beq
\partial_{\alpha} h^{\alpha}_{\; \mu} \, - \, \frac{1}{2}
\partial_{\mu} h^{\alpha}_{\; \alpha} \; = \; 0 \;\; ,
\eeq this perturbation $h_{\mu \nu}$ is related to the source tensor
\beq S_{\mu \nu} \; = \; T_{\mu \nu} \, - \, \frac{1}{2} \, g_{\mu
  \nu} \, T^{\alpha}_{\; \alpha} \;\; .  \eeq through \beq h_{\mu \nu}
\left( \vec{r} \right) \; = \; - \, 4 \, G \; {\displaystyle \int} \,
  {\displaystyle \frac{S_{\mu \nu} \left( \vec{r}\,' \right)} {| \,
      \vec{r}\,' \, - \, \vec{r} \, |}} \, d^3 \vec{r}\,' \;\; .  
\eeq
The stress--energy tensor is denoted by $T_{\mu \nu}$. 
The gravitational potential -- which in the quasi--Newtonian
approximation is nothing but $h_{00} / 2$ -- is sourced by $S_{00}$.
Note that in the case of pressureless matter, one would have $2 S_{00} = T_0^0$
and the gravitational potential would just be sourced by the energy density.
So, the scalar field generates the same gravitational potential
as an equivalent cold matter component with energy density $2
S_{00}$. For this reason, the later quantity can be called the {\it 
effective} density.
In the simplest
case of the neutral scalar field $\Phi$ of Eq.~(\ref{scalar_neutral}),
it reads:
\beq 
\rho_{\rm
    eff} \equiv 2 S_{00} \; = \; 2 \left( \partial_{0} \Phi
\right)^{2} \, - \, 2 V \;\; .  
\eeq 
Note that unlike the true energy
density, $\rho_{\rm eff}$ contains no space--derivatives $\partial_{i}
\Phi$.

In order to generate the gravitational potential well that is observed
inside galaxies, a distribution of dark matter is generally introduced
in addition to the baryon population. An excess of binding ensues
and matter is tied more closely.
Should the scalar field $\Phi$
be responsible for the haloes of galaxies, its effective
density $\rho_{\rm eff}$
would play the role of the ordinary cold dark matter density, and should
in particular be positive.
Since the gravitational potential in a galaxy is essentially static,
we would have to assume -- a priori -- that the time derivative
$\dot{\Phi} = \partial_{0} \Phi$ vanishes. The effective density
$\rho_{\rm eff}$ would therefore reduce to $- \, 2 \, V \left( \Phi
\right)$ and the field potential would have to be negative. This is
actually the solution suggested by \cite{mexico}. However, even if we
are not aware of a principle that strictly forbids a negative
potential, we would prefer to avoid such an unusual assumption.
Nothing would prevent the system from being unstable in that case and
we will therefore disregard
this option.
On the other hand, a static field with positive potential $V$ leads
to a positive Newtonian potential and therefore to repulsion. It would
lessen the attraction of ordinary matter and disrupt galaxies. This is
not completely surprising since the same positive $V$ is proposed
to accelerate the expansion of the universe as its contents repel each
other. This property has led to the hasty conclusion that a scalar field
could not bind matter inside galaxies.

\vskip 0.1cm
Let us assume however that the scalar field $\Phi$ varies much
more rapidly than the system in which it is embedded. Our Milky
Way rotates in $\sim$ 200 million years. If the field changes on
a much shorter timescale, the associated effective density would be felt
through its time average
\beq
\rho_{\rm eff} \; = \;
2 \left< \dot{\Phi}^{2} \right> \, - \, 2 \left< V \right> \;\; .
\eeq
The field $\Phi$ may oscillate for instance at the bottom of the
potential well. The pulsation of the corresponding vibrations
is equal to the scalar mass $m$ and may be derived from
the curvature of the potential at its minimum in $\Phi = 0$
\beq
m^{2} \equiv V''(0) \;\; .
\eeq
Actually the field behaves just like an ordinary harmonic oscillator
whenever the pulsation $m$ is much larger than the wavevector
$k$ of the self--gravitating configuration. This translates into the condition
\beq
{\displaystyle \frac{2 \, \pi}{m}} << R \;\; ,
\eeq
where $R$ is the typical length on which $\Phi$ changes appreciably.
For the haloes of galaxies, $R$ is of order a few kpc. Because $\Phi$
varies in time like $\exp \left( - \, i \, m \, t \right)$, the kinetic
and potential energies are related on average by
\beq
\left< V \left( \Phi \right) \right> \; = \;
\left< {\displaystyle \frac{\dot{\Phi}^{2}}{2}} \right> \;\; .
\label{equality}
\eeq
The effective density may now induce a gravitational attraction insofar as
\beq
\rho_{\rm eff} \; = \; 2 \, \left< V \right> > 0 \;\; .
\eeq
Allowing the field $\Phi$ to vibrate quickly has led to an overall
change of sign with respect to the case considered in \cite{mexico}.
Notice furthermore that the associated pressure vanishes as a result
of~(\ref{pressure}) and (\ref{equality}) so that the scalar fluid behaves
just like non--relativistic matter. Such coherent oscillations have
already been considered in the literature, in the case of the axion in
particular -- see also the interesting
discussion of a quintessence field with a late oscillatory stage in 
\cite{sahni}.

\vskip 0.1cm
Another illustration of a fast evolving field is to make it rotate in
some internal space. We may look for configurations where the
dark energy itself -- and not its time--average -- is rigorously static.
A complex field with a uniformly rotating phase features the simplest
realization of that idea
\beq
\Phi \; = \;
{\displaystyle \frac{\sigma(r)}{\sqrt{2}}} \,
e^{\displaystyle - i \, \omega t} \;\; .
\label{rotatingphase} 
\eeq
If the field is non--interacting but has a mass $m$, the associated
effective density obtains from
\beq
{\displaystyle \frac{\rho_{\rm eff}}{2}} \; = \;
S_{0 0} \; = \; 2 \, \dot{\Phi}^{\dagger} \dot{\Phi} \; - \;
U \left( \Phi \right) \;\; ,
\eeq
where the potential $U = m^{2} \, \Phi^{\dagger} \, \Phi$.
The pressure of the scalar fluid may be approximated by
\beq
P \simeq \dot{\Phi}^{\dagger} \dot{\Phi} \, - \,
U \left( \Phi \right) \;\; ,
\eeq
when the space--derivatives of the field are negligible. This amounts
to assume once again that the typical length $R$ of the system way
exceeds $1 / \omega$. Whenever the condition
$m \simeq \omega$ holds, the pressure is vanishingly
small and the scalar fluid behaves as a non--relativistic component.
The associated effective density becomes
$\rho_{\rm eff} = \omega^{2} \sigma^{2} (r)$ with no explicit
dependence on the time. The complete model will be discussed
in the next section where we will consider the possibility of a
boson--star like system extending over a whole galaxy and playing
the role of a dark halo.

\vskip 0.1cm
We conclude this section by pointing out the difficulty to have a
common explanation for both the local dark matter and the
cosmological quintessence in terms of a scalar field. An excess of
gravitational binding on galactic scales requires the condition
\beq
\dot{\Phi}^{2} \geq V \left( \Phi \right)
\label{condition_gdm}
\eeq
to be fulfilled. Conversely, should the overall pressure $P$ be
negative to account for a cosmological constant, the potential
would have to satisfy the inequality
\beq
{\displaystyle \frac{\dot{\Phi}^{2}}{2}} \leq
V \left( \Phi \right) \;\; .
\label{condition_cc}
\eeq
We conclude that the pressure--to--density ratio $w$ must
exceed the value of $- 1/3$ in order for both conditions
to be simultaneously met. Such a range seems to be already
excluded by the measurements of supernovae SNeIa
\cite{supernovae_omegaL}.

\section{The self--gravitating complex and massive scalar field.}
\label{sec:the_model}

Boson stars have been extensively studied in the past -- see for instance
\cite{ruff,tdlee,jetzer,liddle}. For clarity, we will briefly summarize the
main features of self--gravitating bosons, following closely the presentation
of \cite{tdlee}. We are interested in the stable and bounded configurations
of a complex scalar field obeying the action
\beq
{\cal S} \; = \; {\displaystyle \int} \, \sqrt{- g} \; d^{4}x \;
{\cal L} \left\{ \Phi , \partial_{\mu} \Phi \right\} \; = \;
{\displaystyle \int} \, \sqrt{- g} \; d^{4}x \; \left\{
g^{\mu \nu} \, \partial_{\mu} \Phi^{\dagger} \, \partial_{\nu} \Phi
\; - \; U \left( \Phi \right) \right\} \;\; ,
\eeq
where the potential $U$ is invariant under the global symmetry
\beq
\Phi \longrightarrow \Phi' = e^{\displaystyle i \alpha} \, \Phi \;\; .
\eeq
The conservation of the corresponding conserved current is crucial
for the stability of the boson star. Real scalar fields have no stable
bounded configurations. One can show that all spherically symmetric
minimum energy solutions depend on time only through a rotating
phase so that the complex field $\Phi$ may be expressed as in
Eq.~(\ref{rotatingphase}) -- see for instance the appendix in \cite{tdlee}.
In analogy with the hydrogen atom, such solutions correspond to the
energy eigenstates $(n,l=0,m=0)$. We will see later how the discrete
energy levels $n$ of a boson star are associated with different values
of the rotation parameter $\omega$.
The parametrization~(\ref{rotatingphase}) of $\Phi$ is compatible
with a static isotropic metric
\beq
d \tau^{2} \; = \;
e^{\displaystyle 2 u} \, dt^{2} \, - \,
e^{\displaystyle 2 v} \, \left\{
dr^{2} + r^{2} d \theta^{2} + r^{2} \sin^{2} \theta \, d \varphi^{2}
\right\} \;\; ,
\label{metric_isotropic}
\eeq
where $u$ and $v$ depend only on the radius $r$.
The Klein-Gordon equation reads
\beq
e^{\displaystyle - 2 v} \, \left\{ \sigma'' \, + \,
\left( u' + v' + \frac{2}{r} \right) \, \sigma' \right\} \; + \;
\omega^{2} e^{\displaystyle - 2 u} \, \sigma \; - \;
U'(\sigma) \; = \; 0 \;\; .
\label{Klein_Gordon}
\eeq
The Einstein equations provide two additional independent equations
of  motion. We can choose for instance
\beq
2 v'' \, + \, v'^{2} \, + \, \frac{4 v'}{r} \; = \; - \, 8 \, \pi \, G \;
e^{\displaystyle 2 v} \, \left\{
W \; + \; V \; + \; U \right\} \;\; ,
\label{RG_time}
\eeq
and
\beq
u'' \, + \, v'' \, + \, u'^{2} \, + \, \frac{1}{r} \left( u' + v' \right)
\; = \; 8 \, \pi \, G \; e^{\displaystyle 2 v} \, \left\{
W \; - \; V \; - \; U \right\} \;\; ,
\label{RG_orthoradial}
\eeq
where the kinetic and gradient contributions to the field energy
density are respectively denoted by
\beq
W \; = \; e^{\displaystyle - 2 u} \,
{\displaystyle \frac{\omega^{2} \sigma^{2}}{2}}
\qquad {\rm and} \qquad
V \; = \;
e^{\displaystyle - 2 v} \, {\displaystyle \frac{\sigma'^{2}}{2}}
\;\; .
\eeq
%
From the assumption that boson stars are non singular
configurations, asymptotically flat and of finite energy, severe
restrictions can be put on the boundary conditions for $u$, $v$ and
$\sigma$. In order to avoid any singularity at the origin, the radial
derivatives $u'$, $v'$ and $\sigma'$ must vanish at $r=0$. Actually,
since an angular deficit at the origin would imply an infinite
concentration of energy, we infer the complementary condition
$v(0)=0$. This appears explicitly in the Einstein equations when
they are written in a slightly different way \cite{liddle}. Anyhow,
in this paper, we will focus on the Newtonian regime for which $v$
can be shifted by a constant term without any modification for the
($u$ , $\sigma$) solution.

\vskip 0.1cm
Spacetime is asymptotically Minkowskian if both metric parameters
$u$ and $v$ vanish at infinite distance $r$. More precisely, for
bounded configurations, one expects that on very large distances
the field will appear as a point--like mass $M$ and that
Schwarzschild's metric will be recovered
\beq
e^{\displaystyle u} \; = \; {\displaystyle \frac{r - a}{r + a}}
\quad , \quad
e^{\displaystyle v} \; = \; \left\{
{\displaystyle \frac{r + a}{r}} \right\}^{2}
\quad , \quad
a = GM / 2 \;\; .
\label{Schwartzild}
\eeq
Let us focus now on the finite energy condition. The total energy can
be inferred from the matter and gravitational Lagrangian. The latter
quantity can be calculated by subtracting to the Einstein--Hilbert
action a surface term -- as is usually done for bounded gravitational
objects -- so that
\beq
L_{G} \; = \; {\displaystyle \frac{1}{2 \, G}} \,
{\displaystyle \int_{0}^{\infty}} \, r^{2} dr \,\,
e^{\displaystyle u + v} \, \left\{
v'^{2} \, + \, 2 u' v' \right\} \;\; .
\eeq
The total energy is the sum of the gravitational energy
$E_{G} = - L_{G}$ and of the matter energy
\beq
E_{M} \; = \; {\displaystyle \int_{0}^{\infty}} \, 4 \pi \, r^{2} dr
\,\, e^{\displaystyle u + 3v} \, \left\{
W \, + \, V \, + \, U \right\} \;\; .
\label{energy}
\eeq
Using the Einstein equation (\ref{RG_time}), one can rewrite the total 
energy in terms of the metric and integrate exactly:
\begin{eqnarray}
E_{M} + E_{G} & = &
- \, G^{-1} \, {\displaystyle \int_0^{\infty}} \, r^{2} dr \,\,
e^{\displaystyle u + v} \, \left\{
v'' \, + \, v'^{2} \, + \, \frac{2}{r} v' \, + \, u' v' \right\}
\nonumber \\ 
& & \nonumber \\
& = & - \, G^{-1} \,
\lim_{r \rightarrow\infty} \left\{
r^{2} \, v' \, e^{\displaystyle u + v} \right\} \;\; .
\end{eqnarray}
Inserting the Schwarzschild asymptotic solution~(\ref{Schwartzild}),
one can check that the total mass is the same as the total energy so
that $M = E_{M} + E_{G}$. Following~(\ref{Schwartzild}), the mass is
also the limit of a slighlty different expression 
\begin{eqnarray}
M = - G^{-1} \lim_{r \rightarrow \infty} 
\left\{ r^2 v' e^{v/2} \right\}
\end{eqnarray}
By rewriting this limit as an integral over $r$, and by using the Einstein 
equation (\ref{RG_time}), one is able to express the mass or total energy 
in terms of the field energy density
\beq
E_{M} + E_{G} \; = \;
{\displaystyle \int_{0}^{\infty}} \, 4 \pi \, r^{2} dr \,\,
e^{\displaystyle 5v/2} \, \left\{
W \, + \, V \, + \, U \right\} \;\; .
\eeq
Notice that the gravitational contribution is then contained in
the factor $\exp \left( {\displaystyle 5v/2} \right)$. For bounded
objects, the sum $W+V+U$ should therefore go to zero faster
than $r^{-3}$. 
Another important quantity is the conserved charge associated to the
$U(1)$ global symmetry, i.e., the number of particles minus
antiparticles $N$ 
\beq 
N \equiv {\displaystyle \int_{0}^{\infty}} \, 4
\pi \, r^{2} dr \, \sqrt{- g} \, g^{0 \mu} \, \left\{ i \, \left(
\Phi^{\dagger} \, \partial_{\mu} \Phi \, - \, \partial_{\mu}
\Phi^{\dagger} \, \Phi \right) \right\} \; = \; \frac{2}{\omega} \,
{\displaystyle \int_{0}^{\infty}} \, 4 \pi \, r^{2} dr \,
e^{\displaystyle u + 3v} \, W \;\; .  
\eeq

\vskip 0.1cm The simplest realization of this system occurs with a
quadratic potential $U= m^{2} \, \Phi^{\dagger} \, \Phi$. By
inspecting the Klein--Gordon equation at large radii, one finds that
finite energy solutions may exist only if $m > \omega$\footnote{In the
  opposite case $m < \omega$, the field oscillates at large distance
  like $r^{-1} \sin((\omega^2 - m^2)^{1/2} r)$. It fills the Universe
  with an infinite amount of energy, unless some truncation mechanism
  is put by hand.  This problem arises in particular when $m=0$
  \cite{schunck}, but not for the solutions considered here.}.
Moreover, as soon as $u$ and $v$ -- respectively $u'$ and $v'$ -- are
small with respect to unity -- respectively $1/r$ -- the field
asymptotically behaves as
\beq \sigma \, \propto \, r^{-1} \, \exp
\left\{ - \, \left( m^2 - \omega^{2} \right)^{1/2} r \right\} \;\; .
\eeq 
Dimensionless equations are obtained by rescaling the field by the
Planck mass and the radial coordinate by $m^{-1}$ -- which is
essentially the Compton wavelength of $\Phi$ \beq \bar{\sigma} \, = \,
\sqrt{2 \pi G} \; \sigma \qquad , \qquad \bar{r} \, = \, r \, m \;\; .
\label{definition_sigma_r}
\eeq
Because of the symmetries of the action, the particle mass $m$,
the rotation velocity $\omega$ and the lapse function
$e^{\displaystyle u}$ appear in the dimensionless equations only
through the particular combination
$( \omega / m )^{2} e^{\displaystyle - 2 u}$. It is then convenient
to define the rescaled lapse function
\begin{eqnarray}
e^{\displaystyle - 2 \bar{u}} \; = \; \frac{\omega^{2}}{m^{2}}
\, e^{\displaystyle - 2 u} \;\; .
\end{eqnarray}
Asymptotic flatness imposes a relation between $(\omega/m)$ and
the value of $\bar{u}$ at infinity
\beq
\frac{\omega}{m} \, = \, e^{\displaystyle - \bar{u} (\infty)} \;\; .
\label{limitubar}
\eeq
The solutions can be calculated by integrating a simple system of
three variables -- $\bar{\sigma}$, $\bar{u}$ and $v$ -- from zero to
infinity.  For a given $\bar{\sigma}(0)$, with the assumption that
$v(0) = 0$ and all first derivatives vanish at the origin, there is only
one free boundary condition left, namely the value of $\bar{u} (0)$.
Using an overshooting method, one finds a discrete set of values
$\bar{u} (0)_{n}$ -- with $n=0,...,\infty$ -- such that $\bar{u}$
converges at infinity with $\bar{\sigma}$ and $v$ smoothly
decreasing towards zero. The resulting configurations are the
energy eigenstates of the system. The state with minimal energy is
characterized by the absence of nodes -- of spheres where
$\sigma(r)=0$ -- while each $n$--excited state has got $n$ nodes.

\vskip 0.1cm
Since we will assume that bosons play the role of galactic dark
matter, we only need to study the Newtonian regime in which
$|u|$ and $|v| \ll 1$. In this limit, the system has got additional
symmetries which facilitate the description and classification of
the exact numerical solutions.
The global order of magnitude of $u$ and $v$ depends on the
parameter $\xi$ defined by
$\xi^{2} = 1 - \omega^{2} / m^{2}$, with $\xi \ll 1$ corresponding
to the Newtonian limit. Indeed, one can show \cite{tdlee} that
$u = {\cal O} (\xi^{2})$ and
$v = {\cal O} (\xi^{2})$ while $u + v = {\cal O} (\xi^{4})$.
So, at order ${\cal O} (\xi^{2})$, $u = - v$ -- as usual in the first--order
post--Newtonian approximation \cite{weinberg} -- and the system
follows a simple pair of equations
\begin{eqnarray}
\bar{u}'' \, + \, \frac{2}{\bar{r}} \, \bar{u}' & = &
2 \, \bar{\sigma}^{2} \;\; ,
\label{pure_boson_PN_u}
\\
\bar{\sigma}'' \, + \, \frac{2}{\bar{r}} \, \bar{\sigma}'
& = & 2 \, \bar{u} \, \bar{\sigma} \;\; .
\label{pure_boson_PN_sigma}
\end{eqnarray}
The solutions are therefore left invariant by the following rescaling
\begin{eqnarray}
\bar{u} &\longrightarrow& k ~\bar{u} \;\; , \nonumber \\
\bar{\sigma} &\longrightarrow& k ~\bar{\sigma} \;\; ,
\label{transfos} \\
\bar{r} &\longrightarrow& k^{-1/2} ~\bar{r} \;\; . \nonumber
\end{eqnarray}
This means that any configuration is fully described by its number $n$
of nodes and by the field value $\bar{\sigma} (0)$ at the origin.
In other words, Newtonian solutions with the same number of nodes
are related among each other through the rescaling~(\ref{transfos}).
The invariance of the solution appears more clearly when
relations~(\ref{pure_boson_PN_u}) and (\ref{pure_boson_PN_sigma})
are expressed in terms of the ratios
$S = \bar{\sigma} (r) / \bar{\sigma} (0)$ and
$\bar{u}_{\rm red} = \bar{u} (r) / \bar{\sigma} (0)$
\begin{eqnarray}
&\bar{u}_{\rm red}'' \, + \,
{\displaystyle \frac{2}{x}} \, \bar{u}_{\rm red}' =
2 \, S^{2} \;\; ,&
\label{pure_boson_PN_u_red}
\\
&S'' \, + \, 
{\displaystyle \frac{2}{x}} \, S'  = 
2 \, \bar{u}_{\rm red} \, S \;\; .&
\label{pure_boson_PN_S}
\end{eqnarray}
The length parameter is now described by $x =
m \, r \, \sqrt{\bar{\sigma} (0)} = \bar{r} \, \sqrt{\bar{\sigma} (0)}$.
We conclude that once the number $n$ of nodes is specified, the value
of $\bar{u}_{\rm red}$ at the origin is unique. So is the field
configuration $S(x)$. The ratio
$\bar{u}_{\rm red} (0) = \bar{u} (0) / \bar{\sigma} (0)$ has been
computed for the fundamental state, the first excited states and also in
the limit where $n \rightarrow \infty$. Our results are quoted in
Table~\ref{table1} and are in good agreement with \cite{tdlee}.
%
\begin{table}[h!]
\[
\begin{array}{|c|c|c|c|c|}
\hline
{\rule{0cm}{0.5cm}} {\quad n \quad} &
{\quad \bar{u}_{\rm red}(0) = \bar{u}(0) / \bar{\sigma}(0) \quad} &
{\quad (1 - \omega/m) / \bar{\sigma}(0) \quad} &
{\quad \bar{M} / \sqrt{\displaystyle \bar{\sigma}(0)} \quad} &
{\quad \bar{N} / \sqrt{\displaystyle \bar{\sigma}(0)} \quad} \\
\hline \hline
{\rule{0cm}{0.4cm}}
0 & - 0.91858 & 0.97894 & 2.4 & 1.2 \\
\hline
{\rule{0cm}{0.4cm}}
1 & - 1.2099 & 0.916 & 5.4 & 2.7 \\
\hline
{\rule{0cm}{0.4cm}}
2 & - 1.3437 & 0.892 & 8.4 & 4.2 \\
\hline
{\rule{0cm}{0.4cm}}
3 & - 1.4282 & 0.877 & 11.4 & 5.7 \\
\hline
{\rule{0cm}{0.4cm}}
5 & - 1.5370 & 0.860 & 17.4 & 8.7 \\
\hline
{\rule{0cm}{0.4cm}}
10 & - 1.6831 & 0.839 & 32 & 16 \\
\hline
{\rule{0cm}{0.4cm}}
{\infty} & - 5. & 4.35 & {\infty} & {\infty} \\
\hline
\hline
\end{array}
\]
\caption{
Scaling factors for $\bar{u}$, $( 1 - \omega / m )$, $\bar{M}$ and
$\bar{N}$ for the fundamental $n = 0$ state and a few $n$--excited
states. These numbers are applicable only in the Newtonian limit which
is reached when all quantities $u$, $v$, $\bar{u}$, $\bar{\sigma}$
and $( 1 - \omega / m )$ are small with respect to unity.}
\label{table1}
\end{table}
%
For each configuration, $\omega / m$ could have been calculated
from Eq.~(\ref{limitubar}) but in practice $\bar{u}$ converges very
slowly. We obtain much more precision by taking into account the
asymptotic Schwarzschild expression~(\ref{Schwartzild}) which
implies that
\beq
e^{\displaystyle u} \, = \, 1 - \, \bar{r} \, u' \, + \,
{\cal O} ( r^{-2} ) \;\; .
\eeq
Noticing that $u' = \bar{u}'$, we find in the Newtonian limit
\beq
1 \, - \, \frac{\omega}{m} \, = \,
\lim_{r \rightarrow \infty}
\left\{ \bar{u} \, - \, \bar{r} \, \bar{u}' \right\} \;\; .
\eeq
We also compute a dimensionless mass parameter
\beq
\bar{M} \, = \, \lim_{r \rightarrow \infty}
\left\{ \bar{r}^{2} \, \bar{u}' \right\} \, = \,
{\displaystyle \frac{m M}{{\rm M}_{\rm P}^{2}}} \;\; ,
\eeq
and a rescaled particle number
\beq
\bar{N} \, = \, {\displaystyle \int_{0}^{\infty}} \,
\bar{\sigma}^{2} \, \bar{r}^{2} \, d\bar{r} \, = \, \frac{1}{2} \,
{\displaystyle \frac{m^{2}}{{\rm M}_{\rm P}^{2}}} \, N \;\; .
\eeq
While $(1 - \omega / m)$ scales as $\bar{\sigma}(0)$,
$\bar{M}$ and $\bar{N}$ scale as $\sqrt{\bar{\sigma}(0)}$ with
factors depending on $n$ that are given in Table~\ref{table1}. 

\begin{figure}[h]
\centering 
\leavevmode
\epsfysize=10.cm
\epsfbox{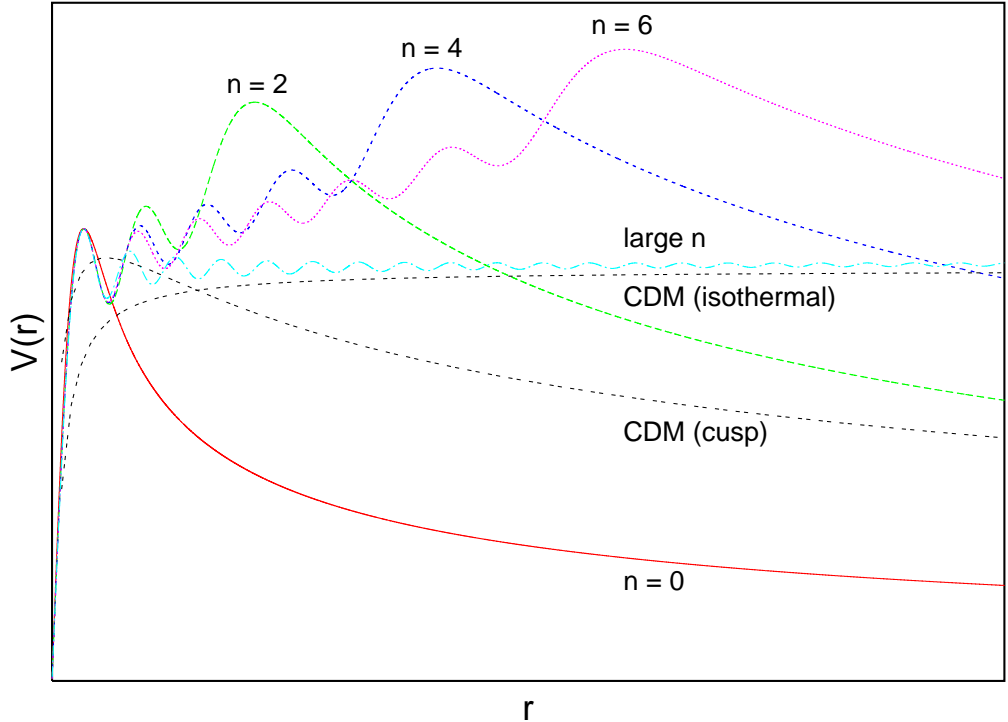}
\vskip 1.cm
\caption{
Rotation curves inside a galactic halo that consists of a pure
self--gravitating scalar field. The corresponding boson is massive but
has no interactions. The fundamental and $n = 2, 4, 6$ states are
featured together with an extremely excited field configuration for
which $n \rightarrow \infty$. Conventional CDM haloes are also
presented for comparison with mass density given by
relations~(\ref{rho_CDM_iso}) -- isothermal distribution -- and
(\ref{rho_CDM_nfw}) -- cuspy profile. Each curve has an arbitrary core
radius and normalization. We choose to normalize the five scalar field
solutions to a common amplitude at the first maximum. These solutions
possess $n+1$ maxima, followed by a decay in $r^{-1/2}$ -- as for any
bounded object. The amplitudes of the first inner maxima are
approximately the same, while the outer ones are bigger. For $n
\rightarrow \infty$, the last maximum and the $r^{-1/2}$ behavior
are rejected far outside the figure, at infinity: we only see a
quasi-flat region with small oscillations.}
\label{fig:fig1}
\end{figure}

\vskip 0.1cm
We now focus on rotation curves inside a toy model of galactic halo
consisting only of bosonic dark matter. Any baryonic contribution
from the disk, the bulge or any other component is neglected here.
Test particles with circular orbits of radius $r$ have rotation speed $v$
\beq
{\displaystyle \frac{v^{2}}{r}} \, = \,
{\displaystyle \frac{\partial}{\partial r}} \, \Phi_{\rm grav}
\, = \, c^{2} u' \;\; .
\eeq
So, the rotation curve is given by $c \, \sqrt{r \, u'}$. In Fig.~\ref{fig:fig1},
we plot this quantity for the fundamental and the $n = 2,4,6$ states as well
as for an extremely excited field configuration with $n \rightarrow \infty$.
We also show the rotation curves associated to the usual fitting functions
for cold dark matter haloes. We first consider an isothermal distribution
with
\beq
\rho_{\rm \, iso} \propto
\left\{ a^{2} \, + \, r^{2} \right\}^{-1} \;\; ,
\label{rho_CDM_iso}
\eeq
where $a$ stands for the core radius. We also feature the case of the
cuspy profile towards which n--body numerical simulations point \cite{nfw}
\beq
\rho_{\rm \, cusp} \propto r^{-1} \, \left\{ a \, + \, r \right\}^{-2} \;\; .
\label{rho_CDM_nfw}
\eeq
In this case, $a$ approximately corresponds to the radius where
peak velocity is reached. The overall normalization and $a$ have been
adjusted in order to match the scalar field halo rotation curves.
%
%
As can be seen on Fig.~\ref{fig:fig1}, the curve associated to the
cold dark matter isothermal distribution~(\ref{rho_CDM_iso}) becomes
flat at large distances as in the large $n$ case. It does not exhibit
however the peak which all the other curves feature. At the outskirts
of the system, the cuspy distribution~(\ref{rho_CDM_nfw}) leads to a
decrease of the rotation velocity with ${\cal V}^{2} \propto \log (r)
/ r$ while, in the $n = 0,2,4,6$ states, the decrease is
typically Keplerian. Near the origin, the scalar field configurations
yield a core with constant density as in the isothermal case. This is
in agreement with recent measurements of the rotation curves of low
surface brightness spirals. Note the small wiggles of the excited
configurations. This could provide an explanation for some peculiar
rotation curves with oscillatory behavior -- as proposed by
\cite{schunck} who obtains similar curves in his massless
model. However we will not consider this possibility in this paper,
since we focus in a first stage on universal rotation curves of spiral
galaxies \cite{persic_salucci_stel}.

\vskip 0.1cm
At this point, we must say a few words about the stability of such
self--interacting bosons. Beside the Newtonian limit, a first
complication arises from the fact that -- for a given number $N$ of
particles and number $n$ of nodes -- there are actually several values
of the mass $M$ corresponding to static, bounded and spherically
symmetric configurations. Only the lowest energy state is stable.
This phenomenon occurs above a critical particle number
$N \sim 0.3 \, ({\rm M}_{\rm P} / m)^{2}$. In the Newtonian
regime, we are much below this scale and, for a given set ($N, n$),
there is a unique static configuration.
A stronger condition is that the gravitating boson system should
be stable against fission into free particles so that $M < N m$.
It was shown in \cite{tdlee} that this criteria is fulfilled by all states in
the Newtonian limit, even for large values of $n$. In that regime,
$M$ tends towards $N \, m$ when $n \rightarrow \infty$ at fixed $N$
(in Table~\ref{table1}, the precision on 
$\bar{M}$ and $\bar{N}$ is not sufficient to see this effect).
Stability against fission is only a necessary condition. More
generally, one should check stability {\bf (i)} at the classical level
under any small perturbation leaving $N$ unchanged and {\bf (ii)} at
the quantum level under tunneling from excited states to the
fundamental state.  The former analysis was performed analytically by
\cite{jetzer} who concludes on the stability of all excited states --
at least in the Newtonian limit. A non--perturbative analysis was
performed numerically in \cite{bala} and yielded the opposite
conclusion. However, as clearly stated in \cite{bala}, these
simulations were based on the most general perturbations which were in
particular allowed to violate the conservation of the particle number
$N$. Therefore, the positive result of \cite{jetzer} seems to apply to
our situation. As far as the stability under tunneling is concerned,
we are not aware of any previous result. Notice anyway that, in order
to be conservative, we will only consider $n = 0$ fundamental state
configurations in what follows.

\section{Comparison with observations.}
\label{sec:rc}

In general, reproducing the observed galactic rotation curves amounts in
modeling the contribution of many components apart from the halo and
the observed luminous disk, like a HI gas, a rotating bar or bulge,
etc. In order to make a strong statement, we will restrict our analysis
to the simplest case of spiral galaxies and on distances smaller than
the optical radius $r_{\rm opt}$ defined as the radius of the sphere
encompassing 83\% of the luminous matter. 
Indeed, for
spiral galaxies and on such distances, the only significant
contributions to the total density arise from a stellar disk with
exponential density profile, plus the unknown dark halo contribution:
one can avoid introducing a plethora of free parameters describing the
other components. On the basis of such considerations, Persic, Salucci
and Stel \cite{persic_salucci_stel} -- hereafter denoted PSS --
performed a detailed statistical
study over about 1100 optical and radio rotation curves. They rescaled each
rotation curve to the same size and amplitude by expressing the radius
as $r / r_{\rm opt}$ and the speed as ${\cal V}(r) / {\cal V}(r_{\rm opt})$.
{\it The rescaled curves were found to depend only on a single parameter, the
luminosity $M_{I}$}. Galaxies were divided in eleven classes depending
on their brightness $M_{I}$ and the authors provided for each group
of spirals the average rotation curve in the range $r < 1.1 \, r_{\rm opt}$.
They showed that for non--luminous galaxies, rotation curves are
increasing near the optical radius while for brighter objects, they tend
to become flat or they even slightly decrease. This result is remarkable
insofar as the dynamical contribution of the luminous disk -- known
up to a constant bias factor $\beta$ -- is slightly decreasing at 
$r_{\rm opt}$.
The immediate conclusion is that faint galaxies are always dominated by
their halo whereas bright spirals only need a very small contribution from
non--luminous matter. This amazing one-to-one correspondance between
the disk and halo core density, depending only on one parameter (the
magnitude), is generally called the disk-halo conspiracy.

\vskip 0.1cm
The purpose of this article is to investigate whether or not a
non--interacting massive scalar field halo may account for the
universal rotation curves of PSS. To achieve this goal, we must
solve once again the Einstein and Klein--Gordon equations,
adding to the former the contribution from the luminous disk.
In order to keep a sherically symmetric metric, we will describe the
gravitational impact of the disk as if it was spherical. This approximation
is reasonable provided that the corresponding contribution to the
mass budget of the system remains small -- which is the case for
faint spirals. In the opposite case, a much more complicated metric needs
to be introduced in order to describe the aspherical distribution of luminous
matter. We therefore add to the time--time Einstein equation the disk
density contribution $\rho$. Should the disk be alone, the rotation curve
would be \cite{persic_salucci_stel}
\beq
{\cal V}^{2}_{\rm disk}(r) \; = \; {\cal V}^{2}_{\rm disk}(r_{\rm opt}) \;
{\displaystyle \frac{1.97 \, (r / r_{\rm opt})^{1.22}}
{\left\{ (r / r_{\rm opt})^{2} + 0.78^{2} \right\}^{1.43}}} \;\; .
\label{v_disk_profile}
\eeq
The density $\rho$ is easily derived from a simple Newtonian calculation
\beq
4 \, \pi \, G \, \rho \, = \,
{\displaystyle \frac{{\cal V}^{2}_{\rm disk}}{r^{2}}} \, + \, 2 \,
{\displaystyle \frac{{\cal V}_{\rm disk}{\cal V}_{\rm disk}'}{r}}
\, = \, {\cal V}_{\rm disk}^{2}(r_{\rm opt}) \;
{\displaystyle \frac{f ( r / r_{\rm opt} )}{r_{\rm opt}^{2}}} \;\; ,
\label{rho_disk}
\eeq
where we introduce the dimensionless function
\beq
f(u) \equiv
{\displaystyle
\frac{4.38 \; u^{-0.78}}{\left\{ u^{2} + 0.78^{2} \right\}^{1.43}}}
\, - \,
{\displaystyle
\frac{5.64 \; u^{1.22}}{\left\{ u^{2} + 0.78^{2} \right\}^{2.43}}} \;\; .
\label{definition_f}
\eeq
Note that this profile is only valid in the range
$0.04 \, r_{\rm opt} < r < 2 \, r_{\rm opt}$ since
relation~(\ref{v_disk_profile}) obtains from the fit of a more
complicated expression involving modified Bessel functions. Below
$r < 0.04 \, r_{\rm opt}$, we maintain a constant density which
would otherwise diverge. We have actually checked that the details
of the disk mass density near the center do not affect our results.

\vskip 0.1cm
Introducing the additional mass distribution $\rho$ of the luminous
disk into the pure scalar case discussed in section~\ref{sec:the_model}
simply amounts to modify the time--time Einstein
relation~(\ref{RG_time}) into
\beq
2 v'' \, + \, v'^{2} \, + \, \frac{4 v'}{\bar{r}} \; = \; - 2 \,
e^{\displaystyle 2 v} \, \left\{
\left( 1 \, + \, e^{\displaystyle - 2 \bar{u}} \right) \, \bar{\sigma}^2
\; + \;  e^{- \displaystyle 2 v} \, \bar{\sigma}^{'2} \; + \;
\bar{\rho} \right\} \;\; ,
\label{RG_time_matter}
\eeq 
where the dimensionless disk density $\bar{\rho}$ may be expressed as
\beq
\bar{\rho} \; = \; {\displaystyle \frac{4 \, \pi \, G}{m^{2}}}
\, \rho \;\; .
\eeq
The radius $\bar{r}$ and the scalar field $\bar{\sigma}$ have been
previously defined in relation~(\ref{definition_sigma_r}). The
other Einstein equation~(\ref{RG_orthoradial}) as well as the
Klein--Gordon relation~(\ref{Klein_Gordon}) are not affected.
It is worth studying the limit of these equations in the Newtonian
regime in order to gain intuition on their scaling behavior. As
before, the relation $u = - v \simeq \bar{u}$ applies in this regime
and the system reduces to
\begin{eqnarray}
\bar{u}'' \, + \, \frac{2}{\bar{r}} \, \bar{u}' & = &
2 \, \bar{\sigma}^{2} \; + \; \bar{\rho} \;\; ,
\label{with_disk_PN_u}
\\
\bar{\sigma}'' \, + \, \frac{2}{\bar{r}} \, \bar{\sigma}'
& = & 2 \, \bar{u} \, \bar{\sigma} \;\; .
\label{with_disk_PN_sigma}
\end{eqnarray}
Relation~(\ref{pure_boson_PN_sigma}) is not modified
whereas the disk mass density $\bar{\rho}$ is introduced
in the right hand side term of expression~(\ref{pure_boson_PN_u})
to yield Eq.~(\ref{with_disk_PN_u}).
As before, the scale invariance of the solution becomes more
obvious when these relations are expressed in terms of the
quantities $S = \bar{\sigma} (r) / \bar{\sigma} (0)$ and
$\bar{u}_{\rm red} = \bar{u} (r) / \bar{\sigma} (0)$. The
system~(\ref{with_disk_PN_u}) and (\ref{with_disk_PN_sigma})
becomes
\begin{eqnarray}
&\bar{u}_{\rm red}'' \,  +  \, 
{\displaystyle \frac{2}{x}} \, \bar{u}_{\rm red}'
 =  2 \, S^{2} \; + \; {\cal R} \;\; , &
\label{with_disk_PN_u_red}
\\
&S'' \,  +  \, 
{\displaystyle \frac{2}{x}} \, S'  = 
2 \, \bar{u}_{\rm red} \, S \;\; .&
\label{with_disk_PN_S}
\end{eqnarray}
The disk density enters through the dimensionless parameter
\beq
{\cal R} \; = \; {\displaystyle \frac{\bar{\rho}}{\bar{\sigma}^{2} (0)}}
\equiv {\displaystyle \frac{4 \, \pi \, G}{m^{2}}} \,
{\displaystyle \frac{\rho}{\bar{\sigma}^{2} (0)}} \;\; .
\eeq
Should ${\cal R}$ vanish, the fundamental $n = 0$ mode
solution for $S(x)$ would be uniquely determined. Taking
advantage of relations~(\ref{rho_disk}) and (\ref{definition_f})
allows to express the disk contribution ${\cal R}$ as
\beq
{\cal R} \; = \;
{\displaystyle
\frac{{\cal V}^{2}_{\rm disk}(r_{\rm opt})}{\bar{\sigma} (0)}}
\; \left\{
x_{\rm opt}^{-2} \;
f \left( {\displaystyle \frac{x}{x_{\rm opt}}} \right) \right\}
\;\; .
\eeq
The general scale--invariant solution $S(x)$ in the presence
of a disk depends now on the parameters
\beq
\alpha \equiv x_{\rm opt} \, = \,
m \, r _{\rm opt} \, \sqrt{\bar{\sigma} (0)}
\quad {\rm and} \quad
\gamma \, = \, {\displaystyle
\frac{{\cal V}^{2}_{\rm disk}(r_{\rm opt})}{\bar{\sigma} (0)}}
\;\; ,
\eeq
which come into play through the disk density
\beq
{\cal R} \left\{ \alpha , \gamma \right\} \; = \;
\gamma \, \alpha^{-2} \, f \left( x / \alpha \right) \;\; .
\eeq
The radial extension $x_{\rm opt}$ of the matter disk relative
to the scalar halo is accounted for by the parameter $\alpha$.
Since the scalar field generates a rotation curve whose magnitude
${\cal V}^{2}$ scales as $\bar{\sigma} (0)$, the quantity $\gamma$
measures the dynamical impact of the disk relative to that of the halo.
These parameters determine therefore the size and the mass of the
disk with respect to the halo in which it is embedded. The actual
scale of the entire system is specified in turns by -- say -- the optical
radius $r_{\rm opt}$ and the disk speed
${\cal V}_{\rm disk}(r_{\rm opt})$. Once the configuration
$\left\{ \alpha , \gamma \right\}$ is chosen, the behavior of the
field -- given by the scale--invariant solution $S_{\alpha \gamma}(x)$
to the system~(\ref{with_disk_PN_u_red}) and (\ref{with_disk_PN_S})
-- is completely determined. The full rotation curve which both disk
and halo generate may be readily derived.

\begin{figure}[h!]
\centering 
\leavevmode
\epsfysize=7.cm
\epsfbox{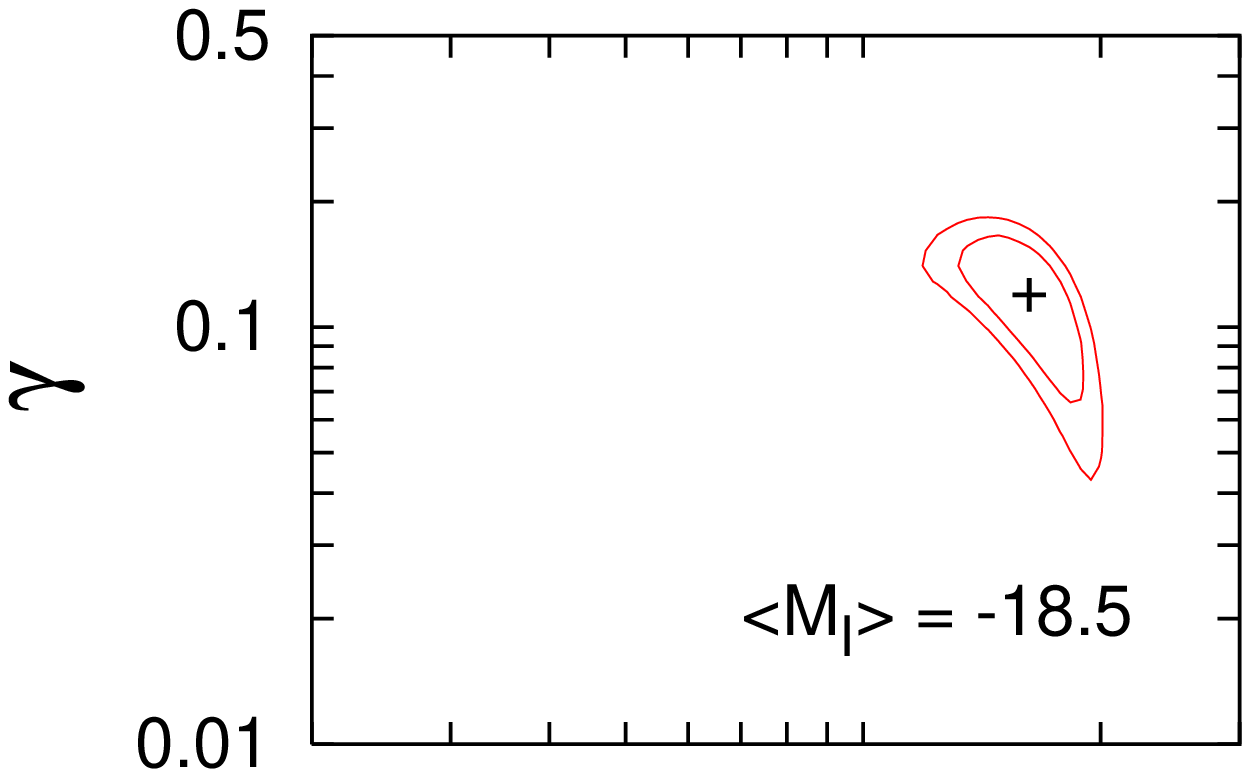}
\hspace{-2.5cm}
\epsfysize=7.cm
\epsfbox{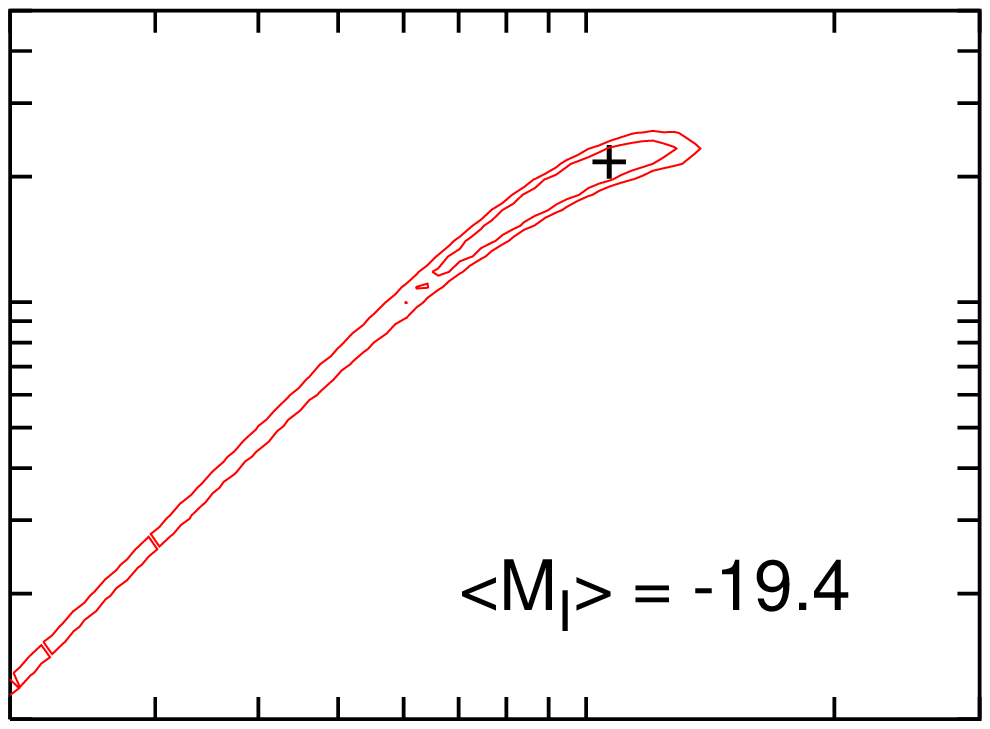}
\vspace{-1.4cm}
\\
\leavevmode
\epsfysize=7.cm
\epsfbox{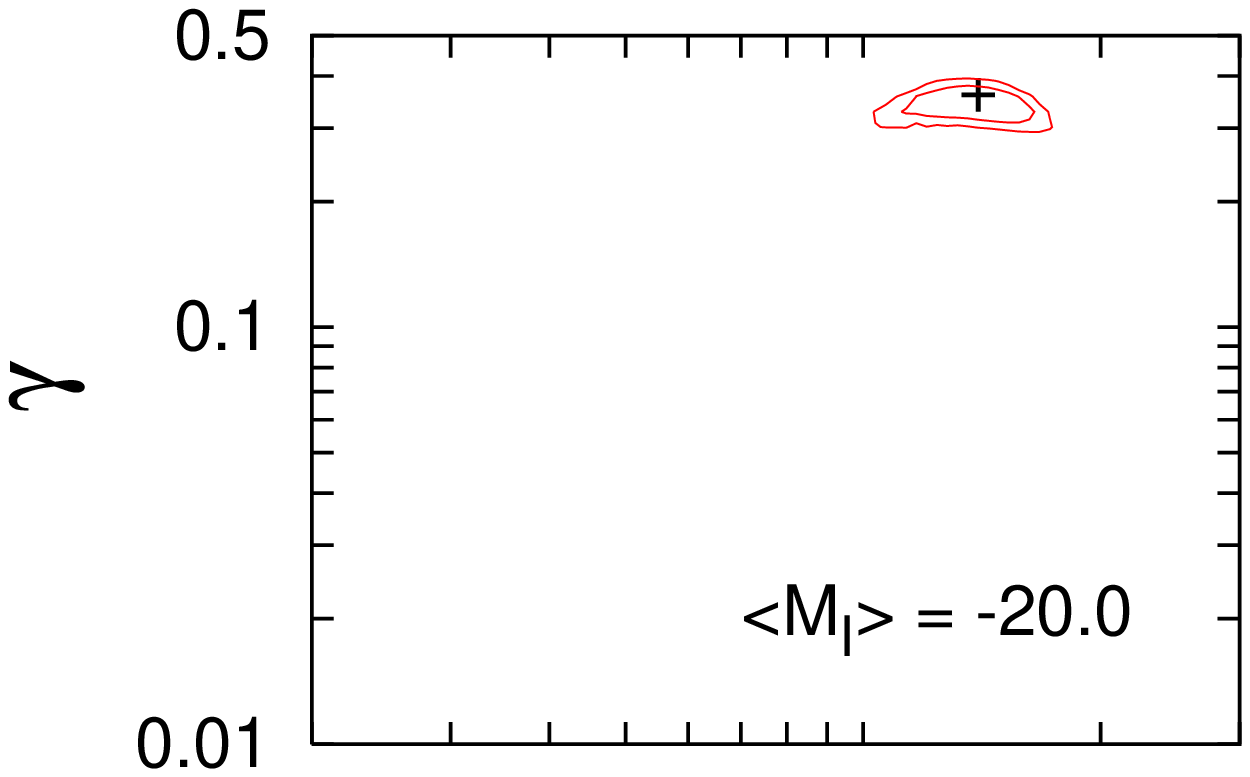}
\hspace{-2.5cm}
\epsfysize=7.cm
\epsfbox{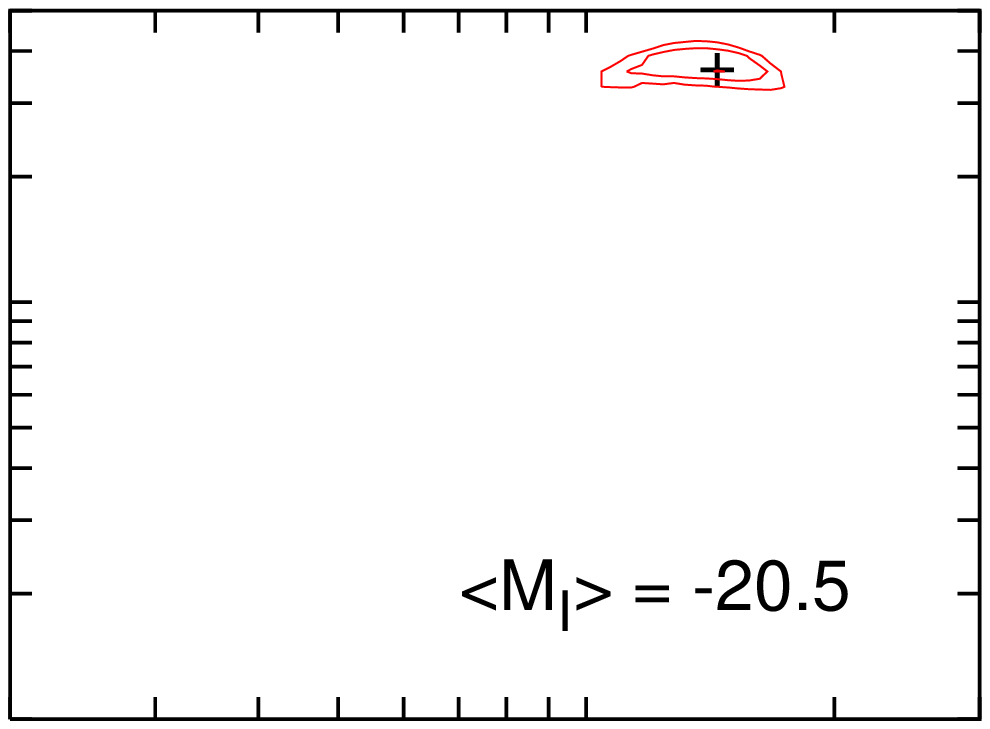}
\vspace{-1.4cm}
\\
\leavevmode
\epsfysize=7.cm
\epsfbox{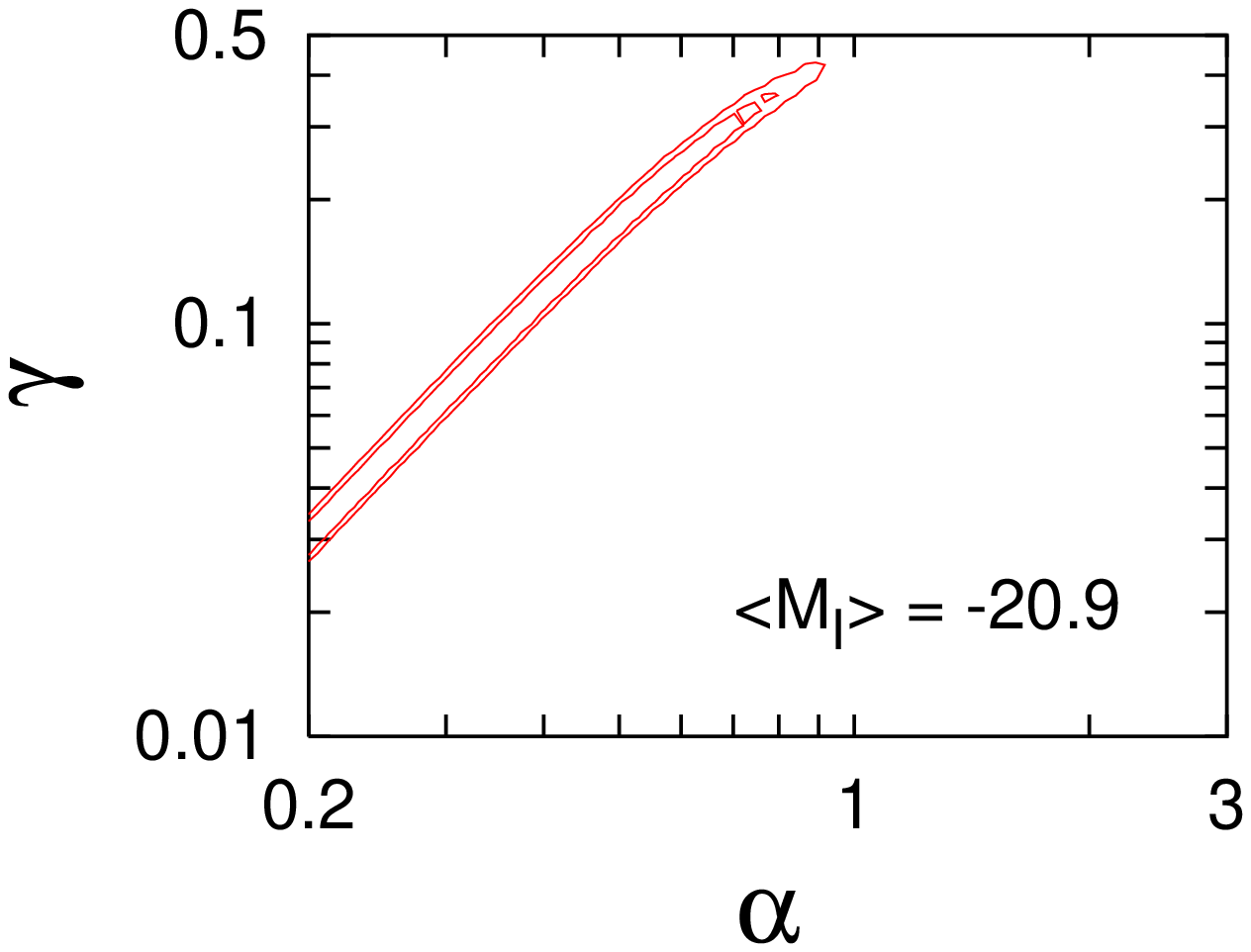}
\hspace{-2.5cm}
\epsfysize=7.cm
\epsfbox{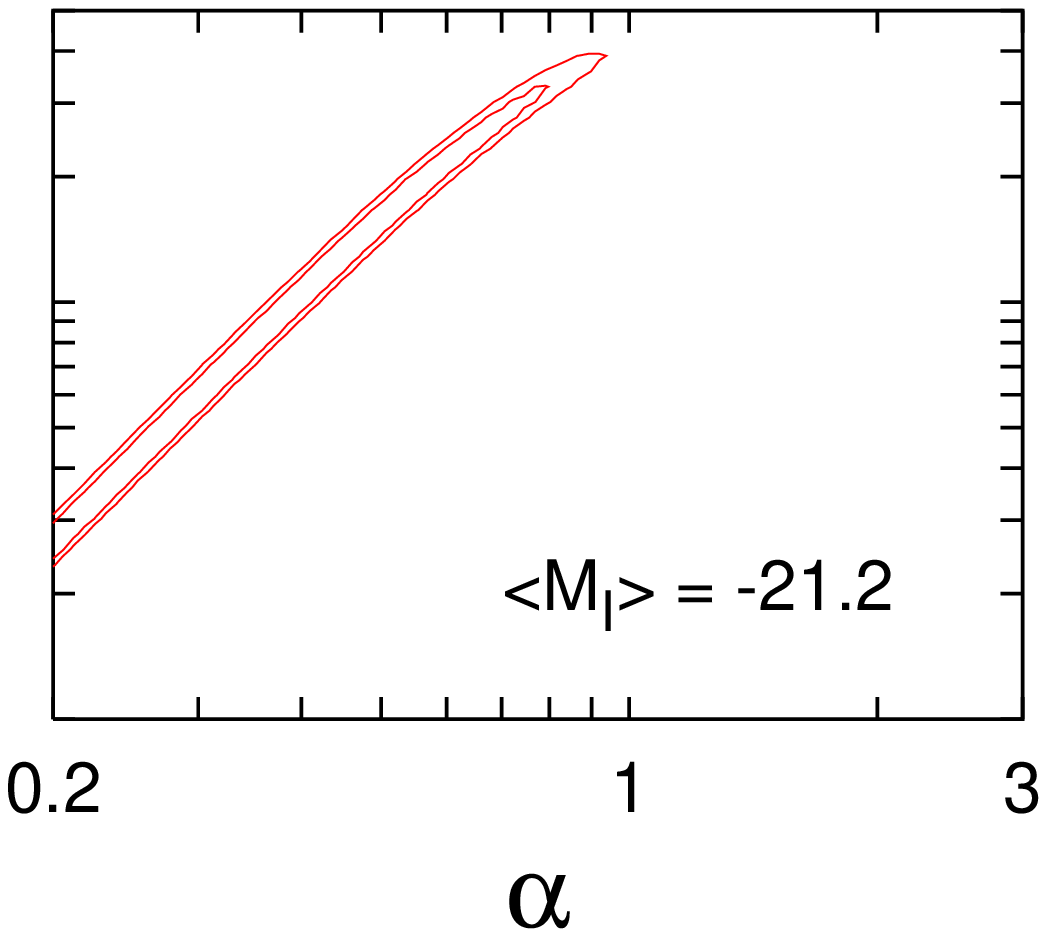}
\vspace{1.cm}
\\
\caption{
  Two--dimensional likelihood contours in the $\{ \alpha, \gamma \}$
  space, for six galaxy classes ordered by growing magnitude.  In the
  four first cases, the crosses show the best--fit models. For the
  last two cases, the minimum is strongly degenerate along a line
  ranging from $\alpha = 0$ to $\alpha = 0.5$. The curves correspond
  to $\chi^{2} - \chi^{2}_{\rm min} = 3.2$ (resp. $6.2$), which would
  give the 68\% (resp. 95\%) allowed region if the experimental errors
  could be rigorously interpreted as 1--$\sigma$ gaussian errors.}
\label{fig:fig2}
\end{figure}

\begin{figure}[h!]
\centering 
\leavevmode
\epsfysize=7.cm
\epsfbox{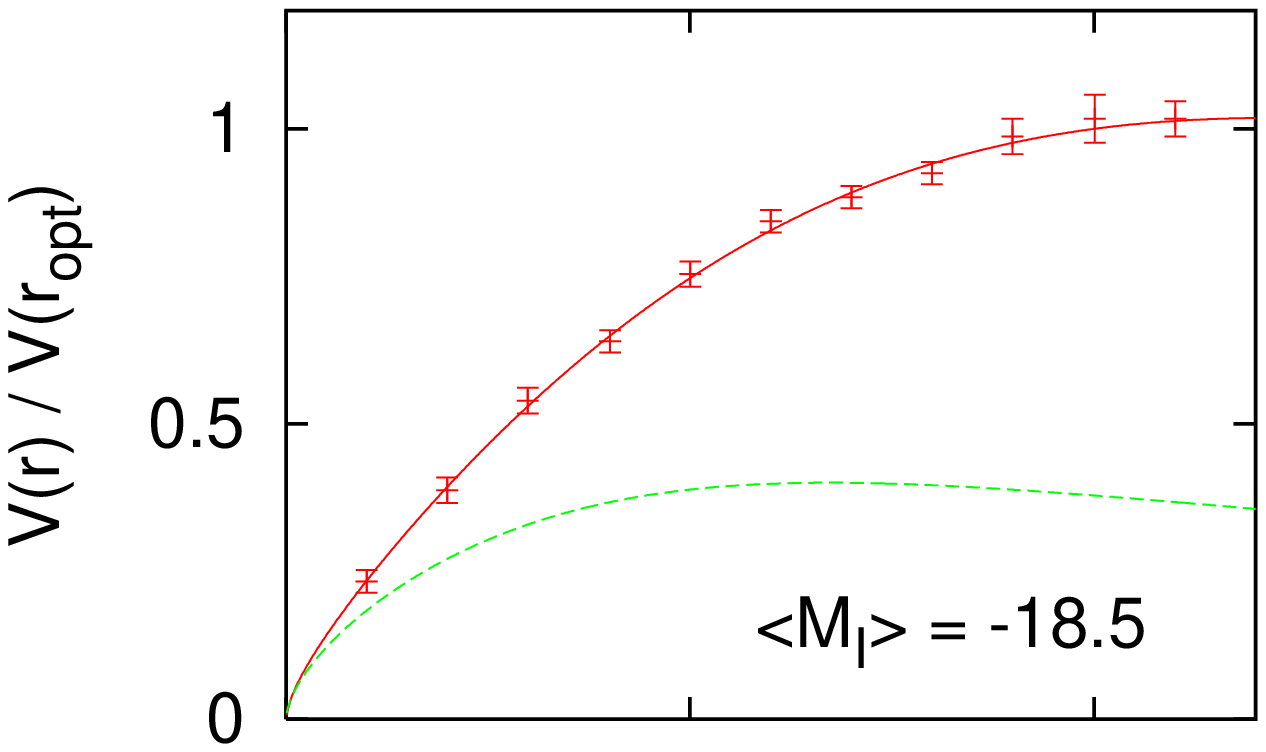}
\hspace{-2.2cm}
\epsfysize=7.cm
\epsfbox{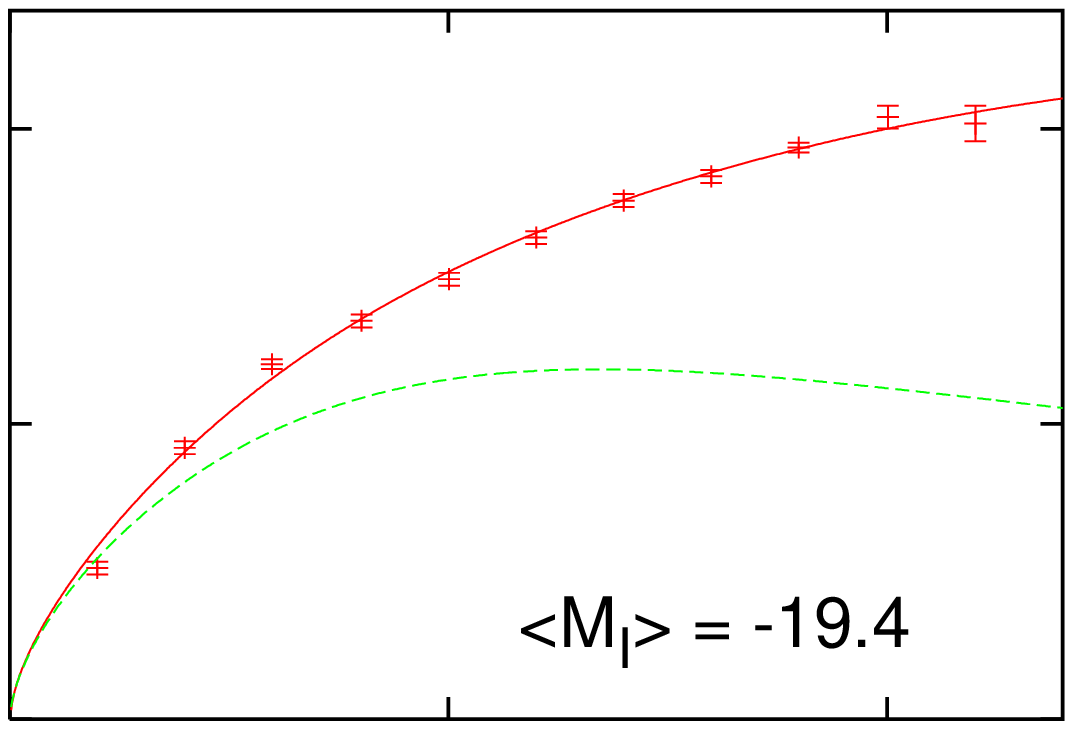}
\vspace{-1.4cm}
\\
\leavevmode
\epsfysize=7.cm
\epsfbox{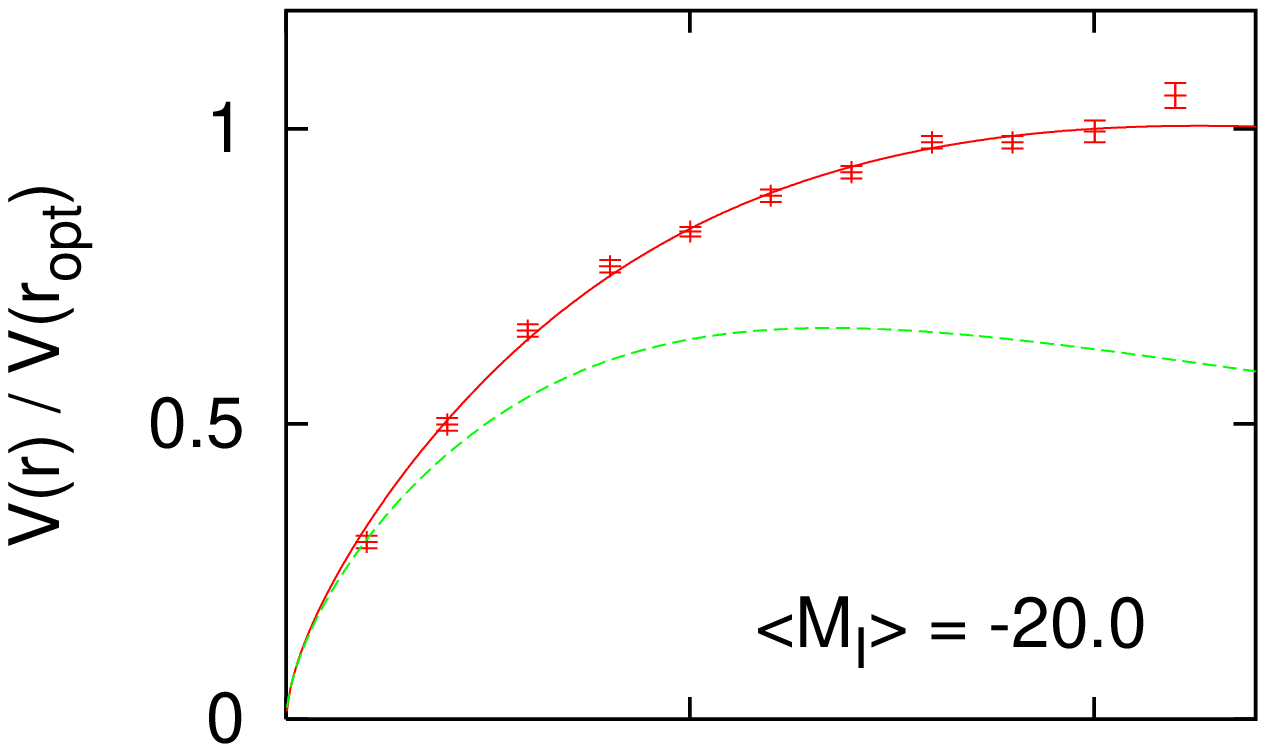}
\hspace{-2.2cm}
\epsfysize=7.cm
\epsfbox{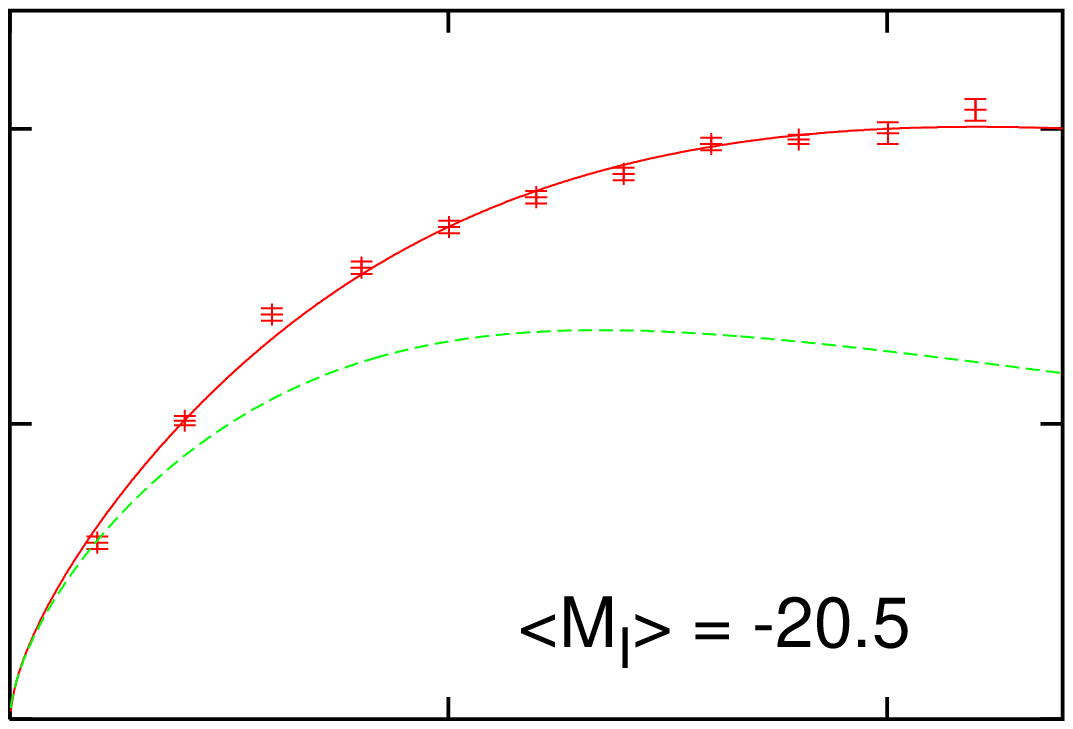}
\vspace{-1.4cm}
\\
\leavevmode
\epsfysize=7.cm
\epsfbox{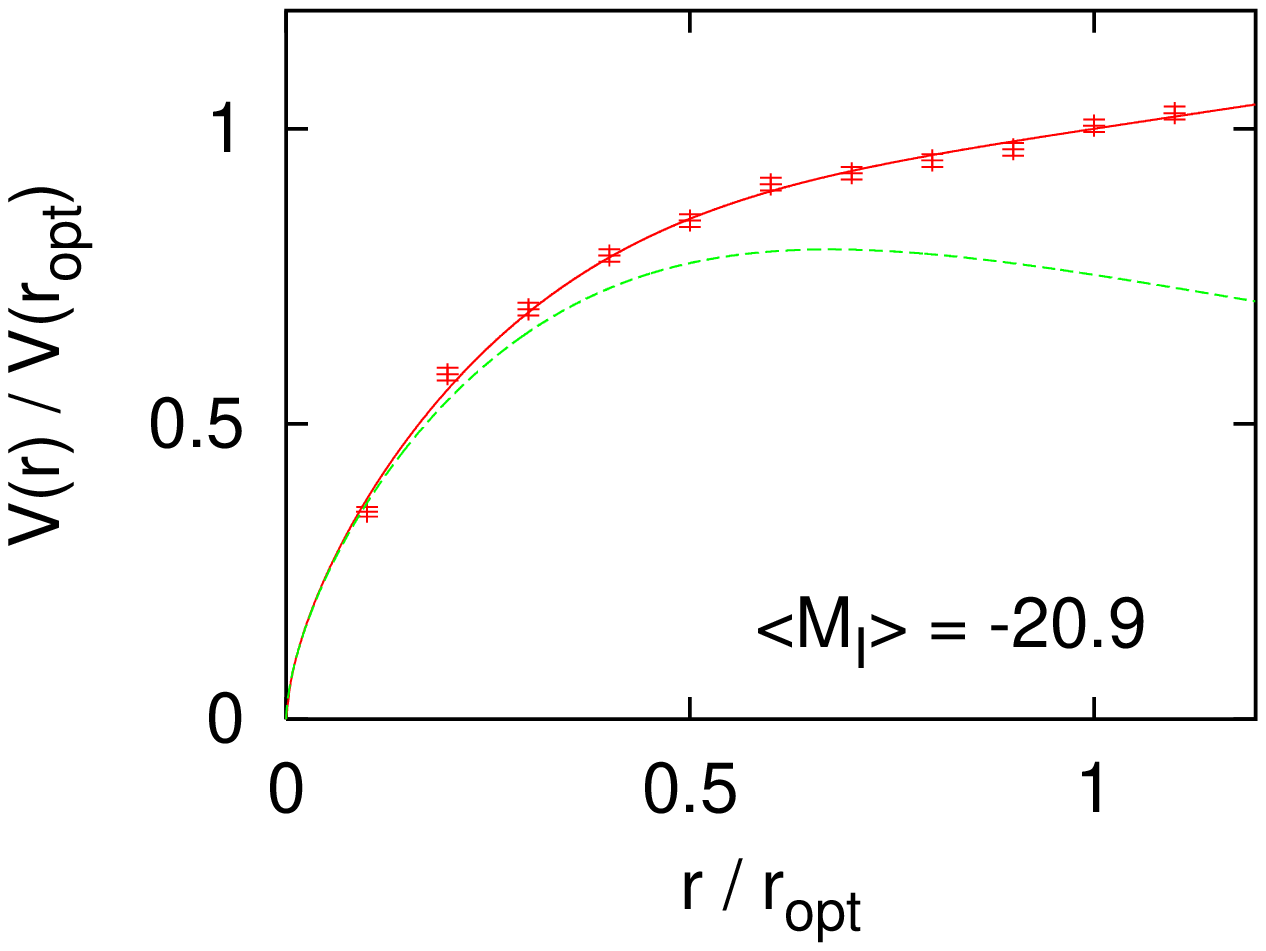}
\hspace{-2.2cm}
\epsfysize=7.cm
\epsfbox{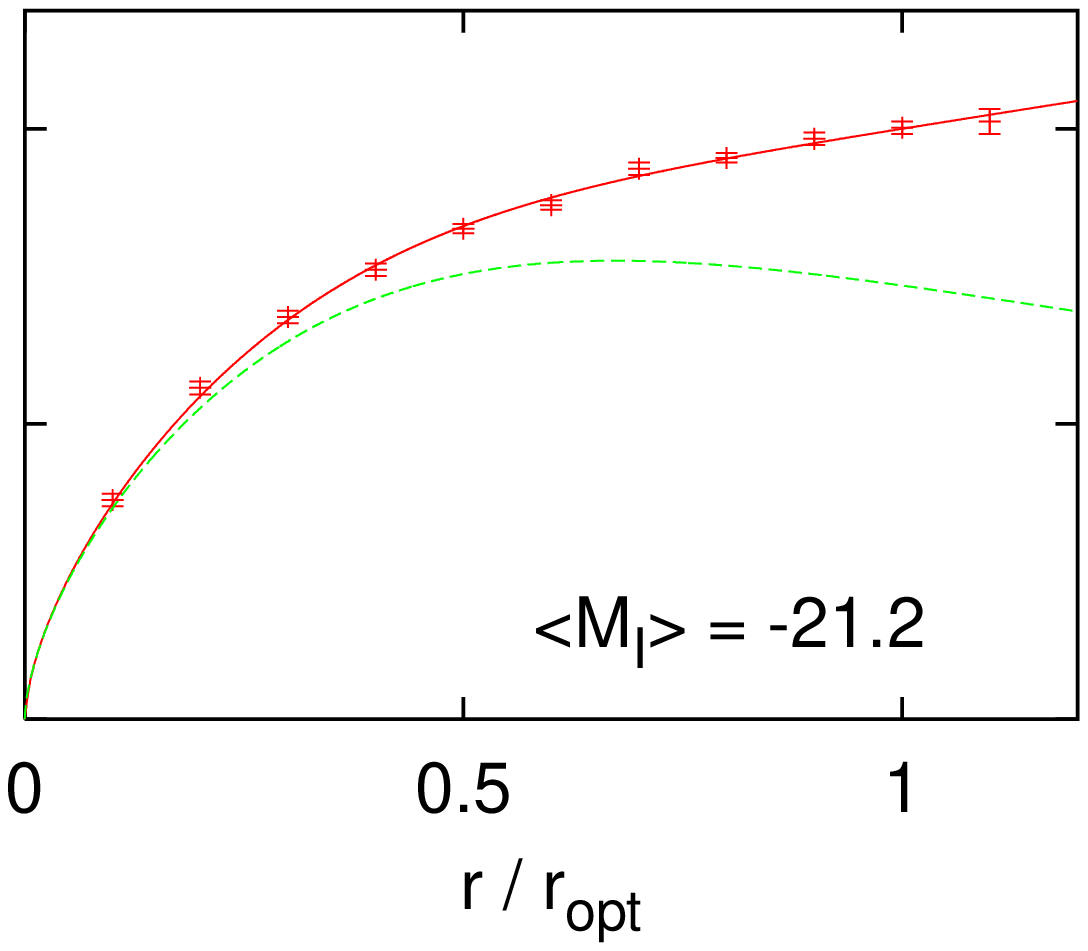}
\vspace{1.cm}
\\
\caption{
  In each panel, the theoretical rotation curve corresponds to the
  $\left\{ \alpha , \gamma \right\}$ configuration that provides the
  best fit to the data. The dashed lines shows the contribution of the
  disk, which increases with the galaxy magnitude as for usual CDM
  models.}
\label{fig:fig3}
\end{figure}

%
\vskip 0.1cm
With a $\chi^{2}$ analysis, we obtain likelihood contours in the
two--dimensional
\footnote{The PSS data include an error bar also at $r = r_{\rm opt}$,
reflecting the cumulated observational uncertainties at the point chosen
for rescaling. So, in each $\chi^{2}$ calculation, we must marginalize
over an overall data normalization factor, which means that we have
three free parameters and $11 - 3 = 8$ degrees of freedom.}
free parameter space $\{ \alpha, \gamma \}$ shown on Fig.~\ref{fig:fig2}.
%
%
We use only the six less luminous galaxy classes from PSS since they
are the most relevant probes of the halo contribution and {\it a priori}
the closest cases to  spherical symmetry -- brighter galaxies require only a
small halo contribution, at least inside the optical radius. The minimum
$\chi^{2}$ values are (2.5, 23.6, 20.9, 29.4, 17.8,7.6), and the
corresponding best--fit rotation curves are shown on Fig.~\ref{fig:fig3}.
%
%
We will not over--interpret the absolute value
of the $\chi^{2}$ in terms of goodness--of--fit, because we do not
know the exact meaning of the data error bars. A careful examination
shows that the points are not distributed according to their very
small errors, at least if the data are to be
explained by smooth curves -- this is visible for instance with the
third point in the $< M_{I} > = - 20.5$ case, which explains why the
minimal $\chi^{2}$ is only 29.4. Therefore, there is a hint
either that the errors are slightly underestimated, or that the data
feature small wiggles corresponding e.g. to spiral arms, that should
enter into a better modelization of the disk density. Anyway, even
with the given error bars, the $\chi^{2}$ are already fairly good and
our scalar halo model seems to fit universal rotation curves at least
as well as the toy--model CDM halo used by PSS.
We also note that for the last two cases, the minimum is strongly
degenerate along a line ranging from
$\alpha = 0$ to $\alpha = 0.5$. So, for the most luminous
galaxies, the data provide only a lower bound on the halo size, while
for the other cases a specific value is preferred.

\begin{figure}[h!]
\centering 
\leavevmode
\epsfysize=7.cm
\epsfbox{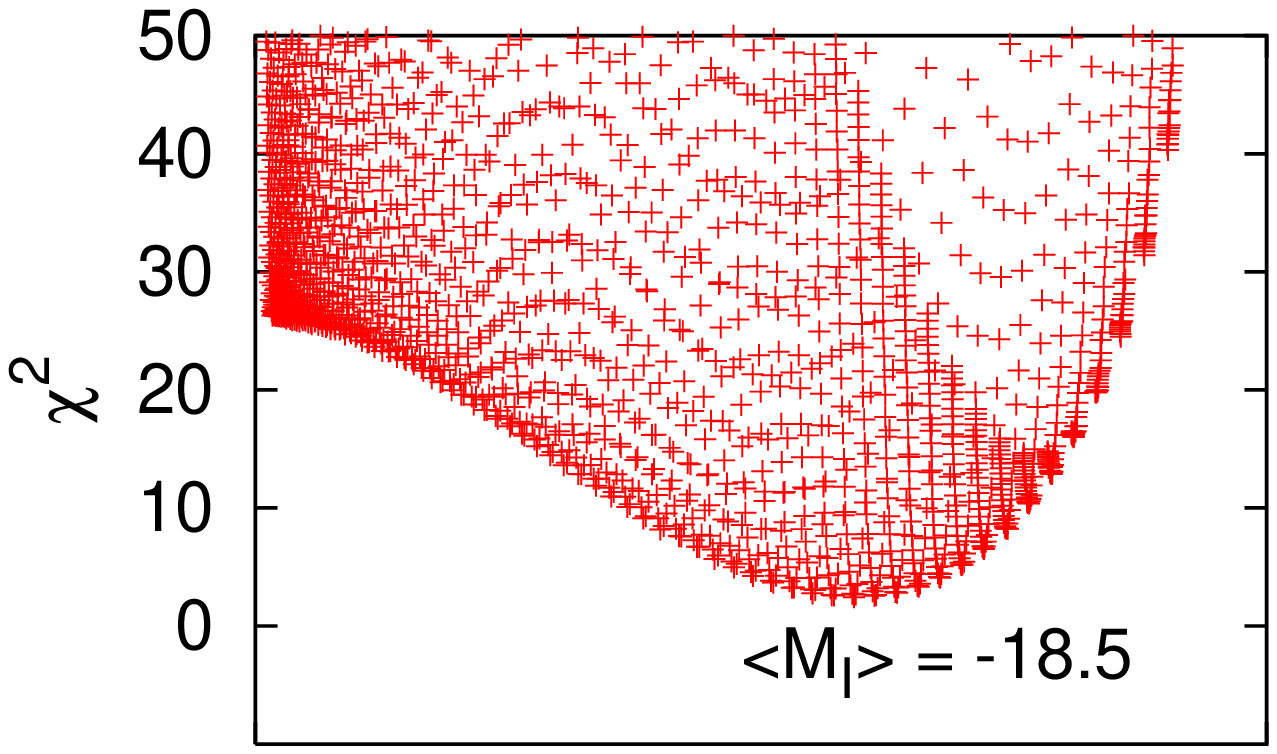}
\hspace{-2.5cm}
\epsfysize=7.cm
\epsfbox{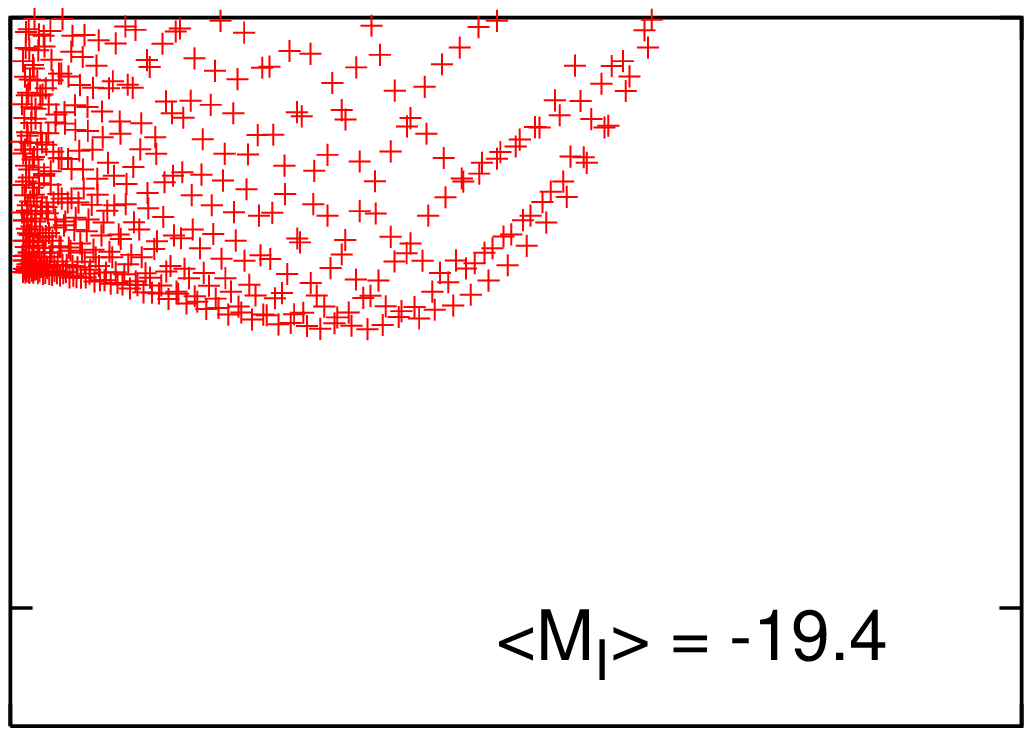}
\vspace{-1.4cm}
\\
\leavevmode
\epsfysize=7.cm
\epsfbox{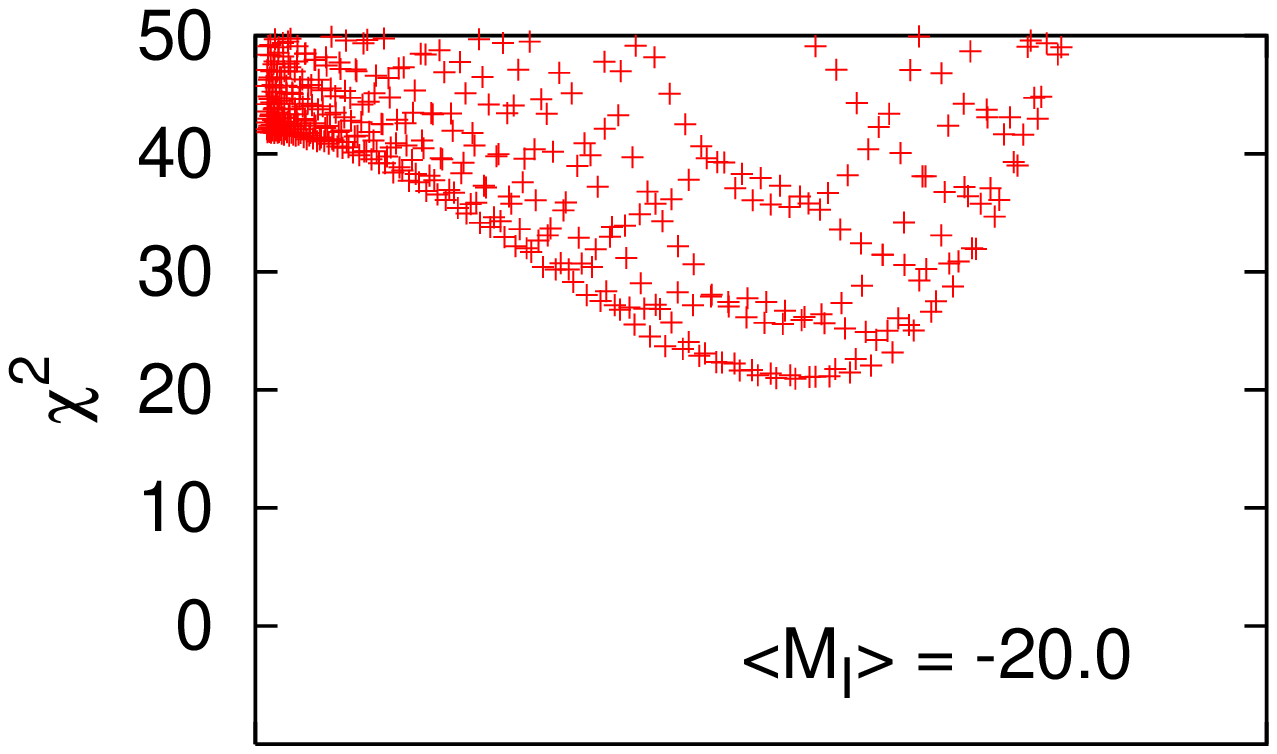}
\hspace{-2.5cm}
\epsfysize=7.cm
\epsfbox{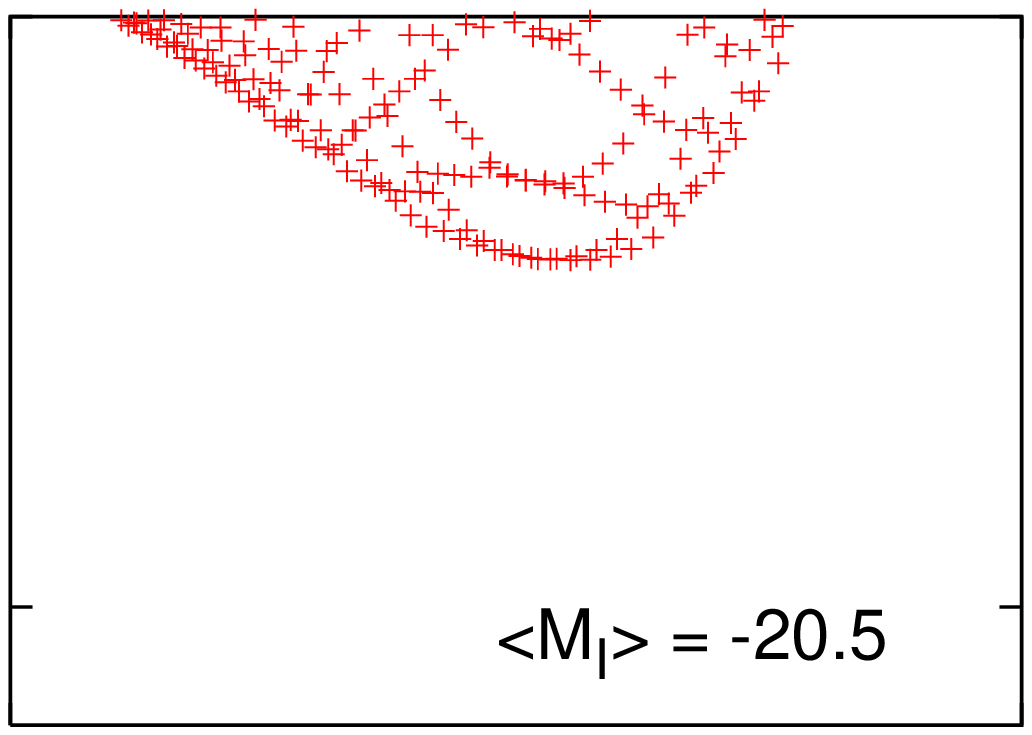}
\vspace{-1.4cm}
\\
\leavevmode
\epsfysize=7.cm
\epsfbox{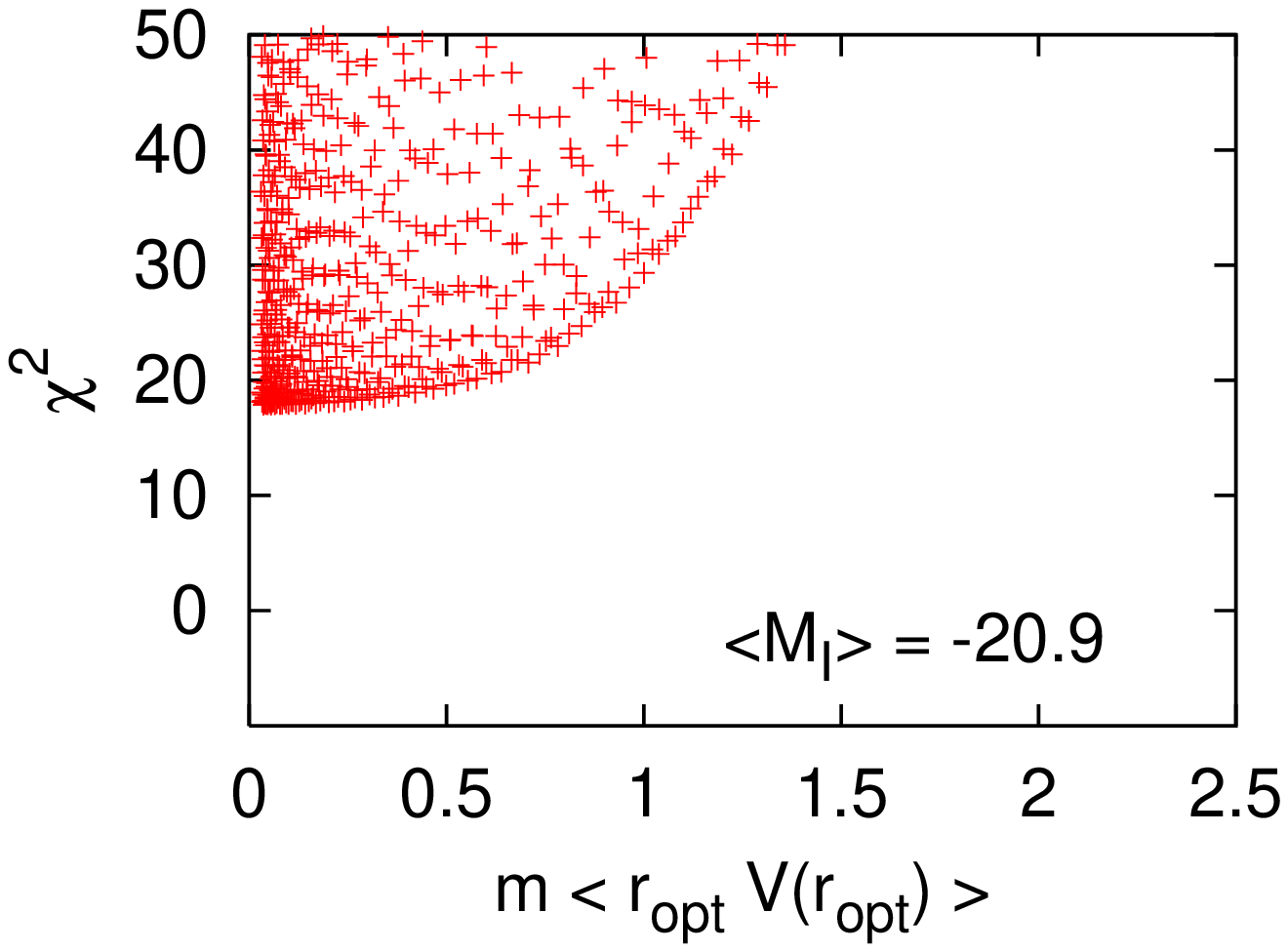}
\hspace{-2.5cm}
\epsfysize=7.cm
\epsfbox{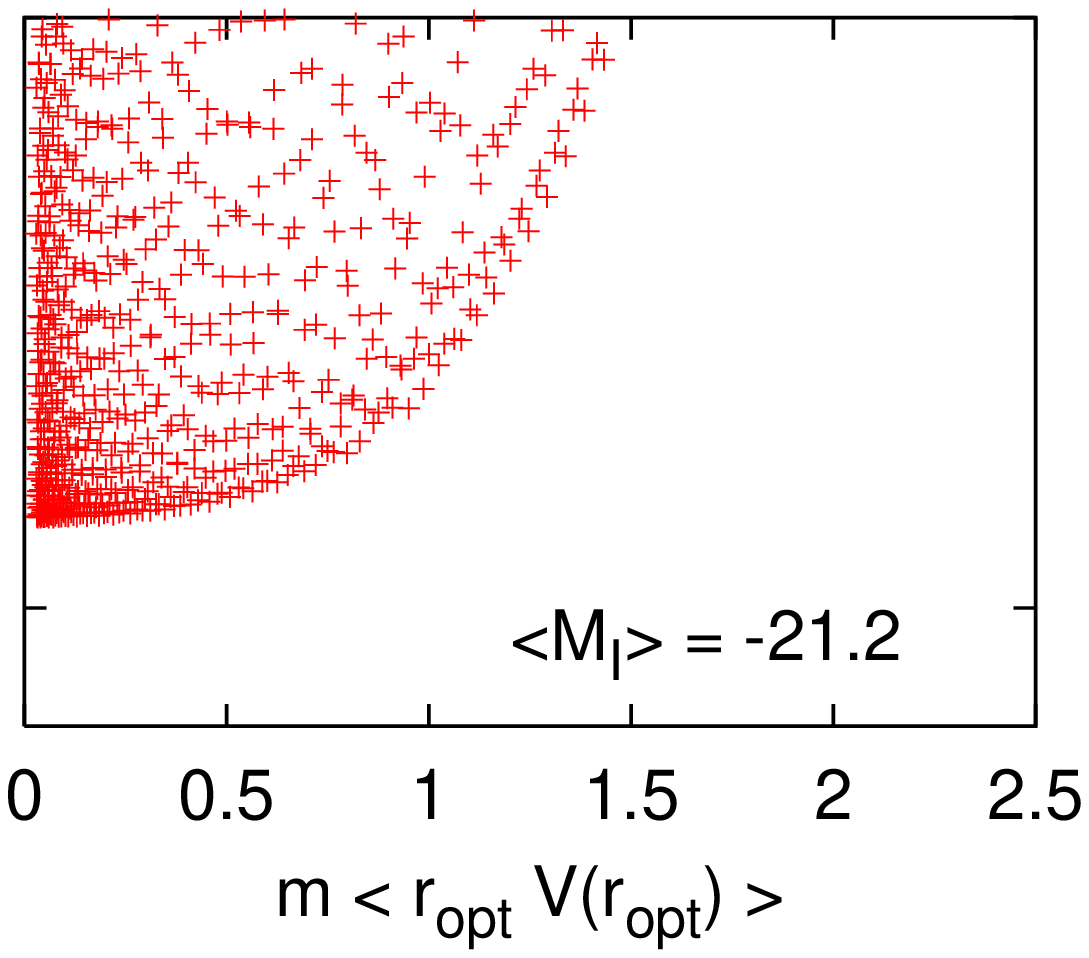}
\vspace{1.cm}
\\
\caption{
The constellation of
$\left\{ \alpha , \gamma \right\}$ configurations is featured in the plane
$(m \left< r_{\rm opt} {\cal V}_{\rm opt} \right> , \chi^{2})$.
Faint galaxies have a preferred mass while the
brighter ones in the two last panels provide only an upper bound.}
\label{fig:fig4}
\end{figure}
%
\begin{figure}[h!]
\centering 
\leavevmode
\epsfysize=7.cm
\epsfbox{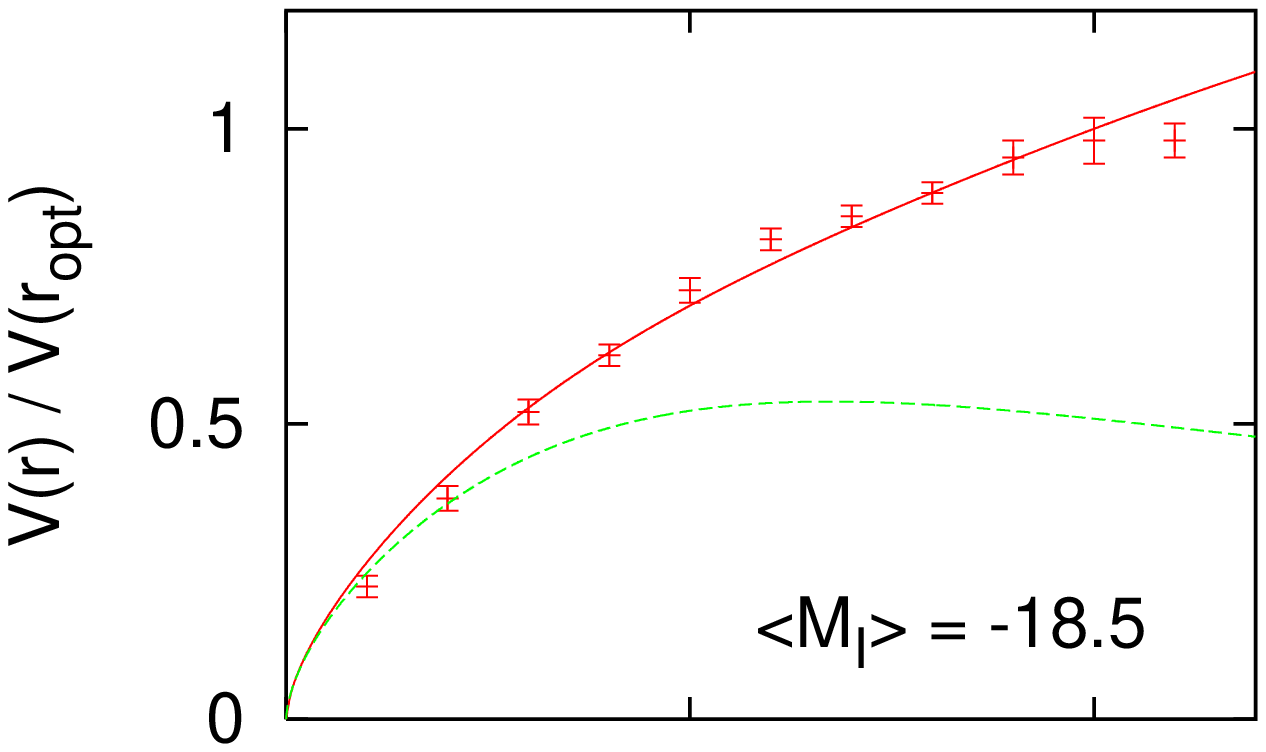}
\hspace{-2.2cm}
\epsfysize=7.cm
\epsfbox{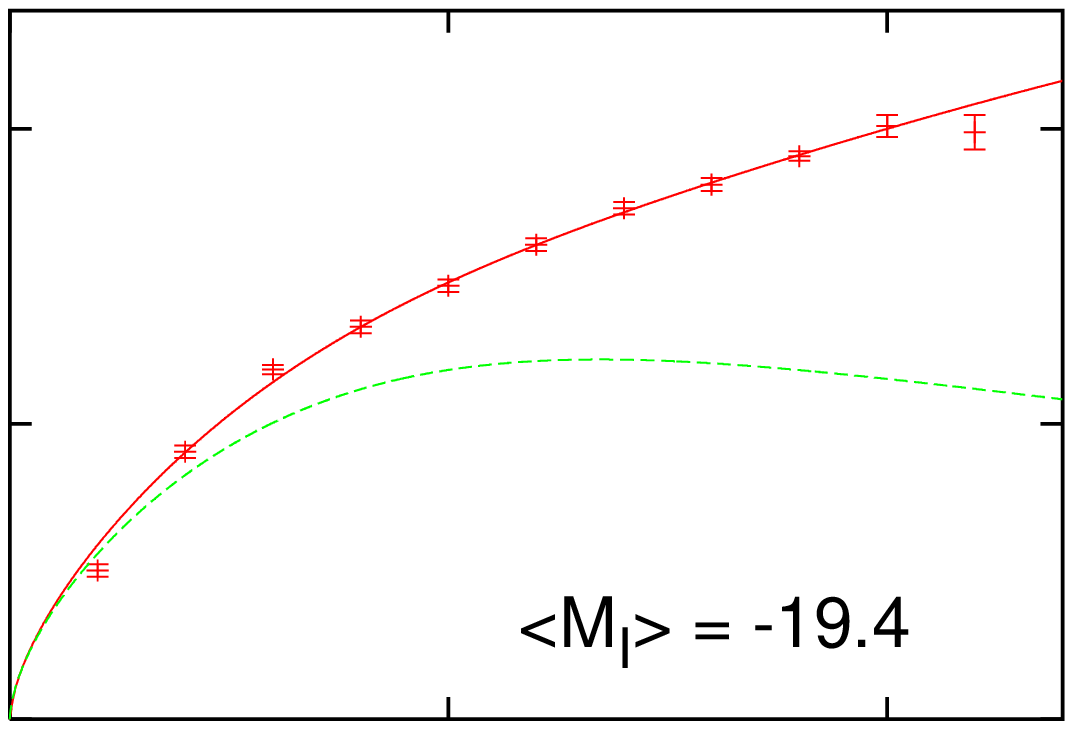}
\vspace{-1.4cm}
\\
\leavevmode
\epsfysize=7.cm
\epsfbox{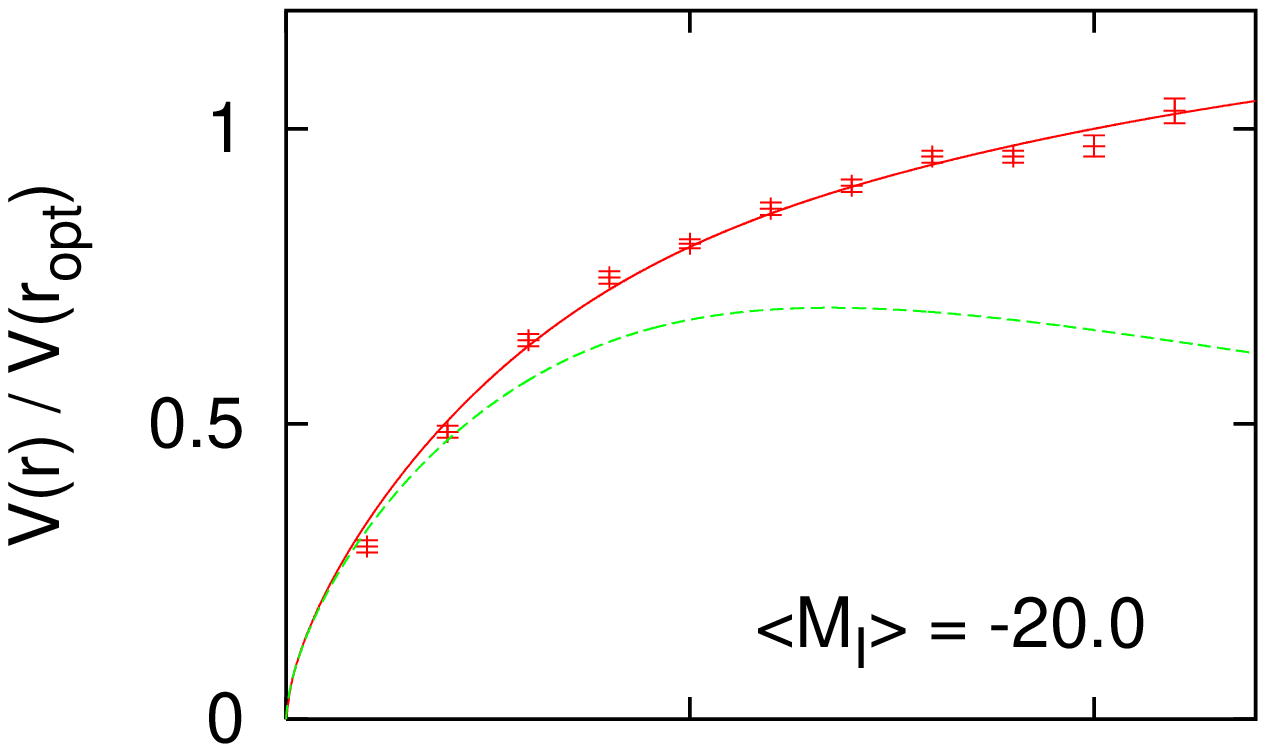}
\hspace{-2.2cm}
\epsfysize=7.cm
\epsfbox{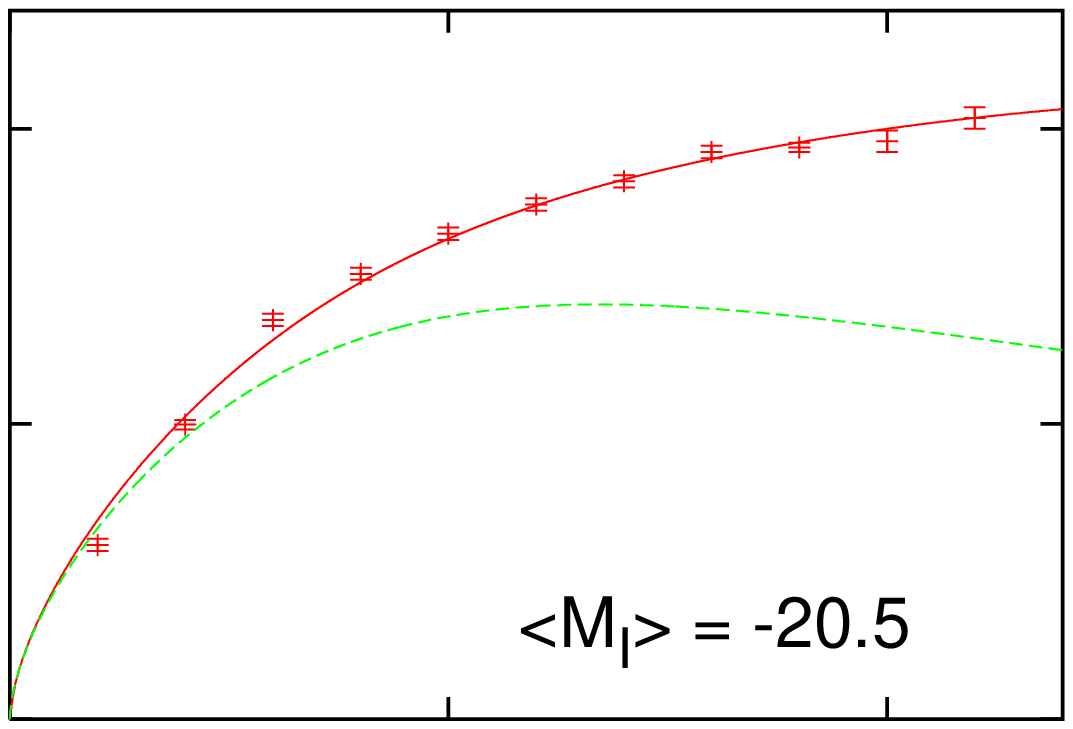}
\vspace{-1.4cm}
\\
\leavevmode
\epsfysize=7.cm
\epsfbox{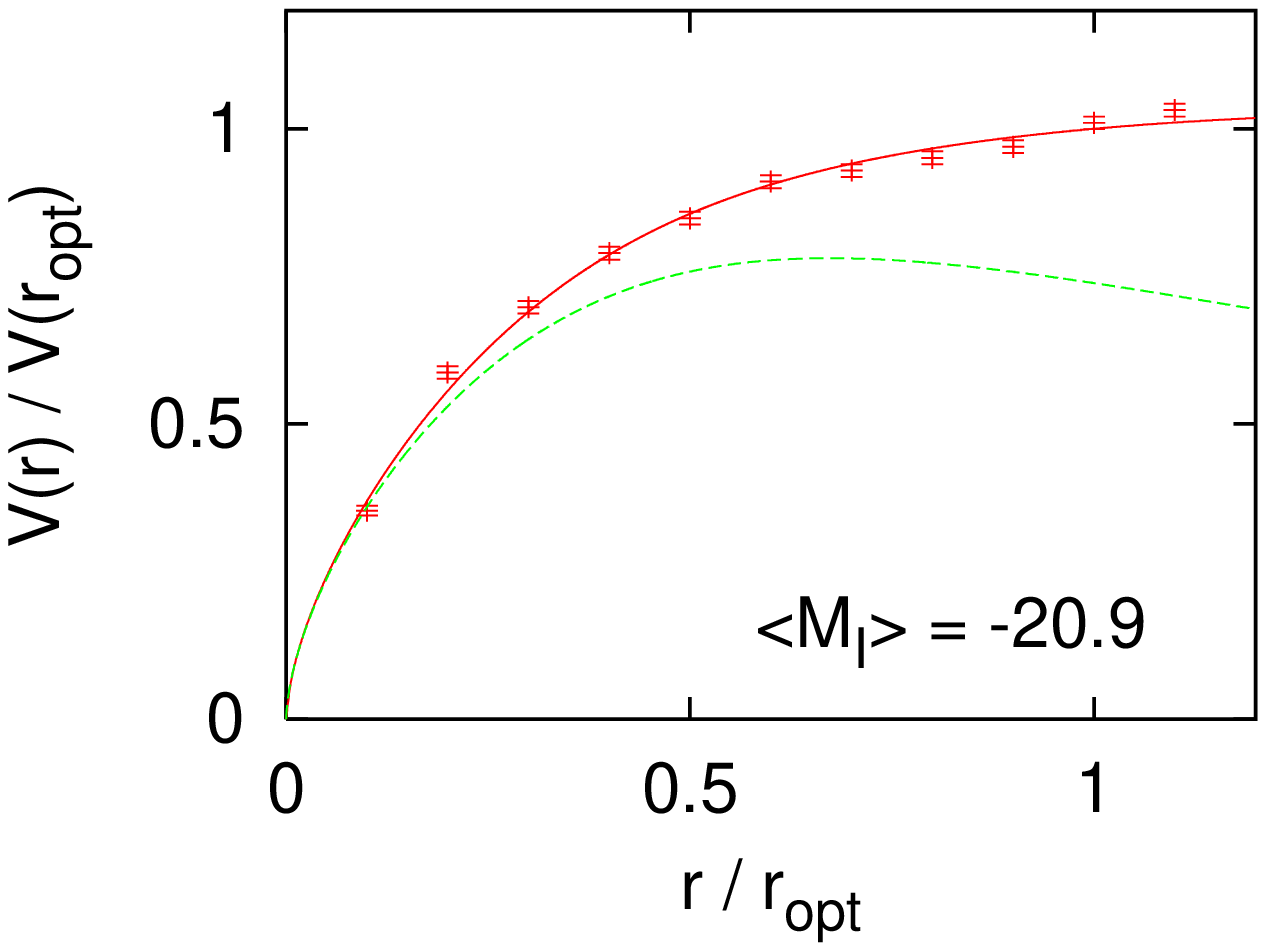}
\hspace{-2.2cm}
\epsfysize=7.cm
\epsfbox{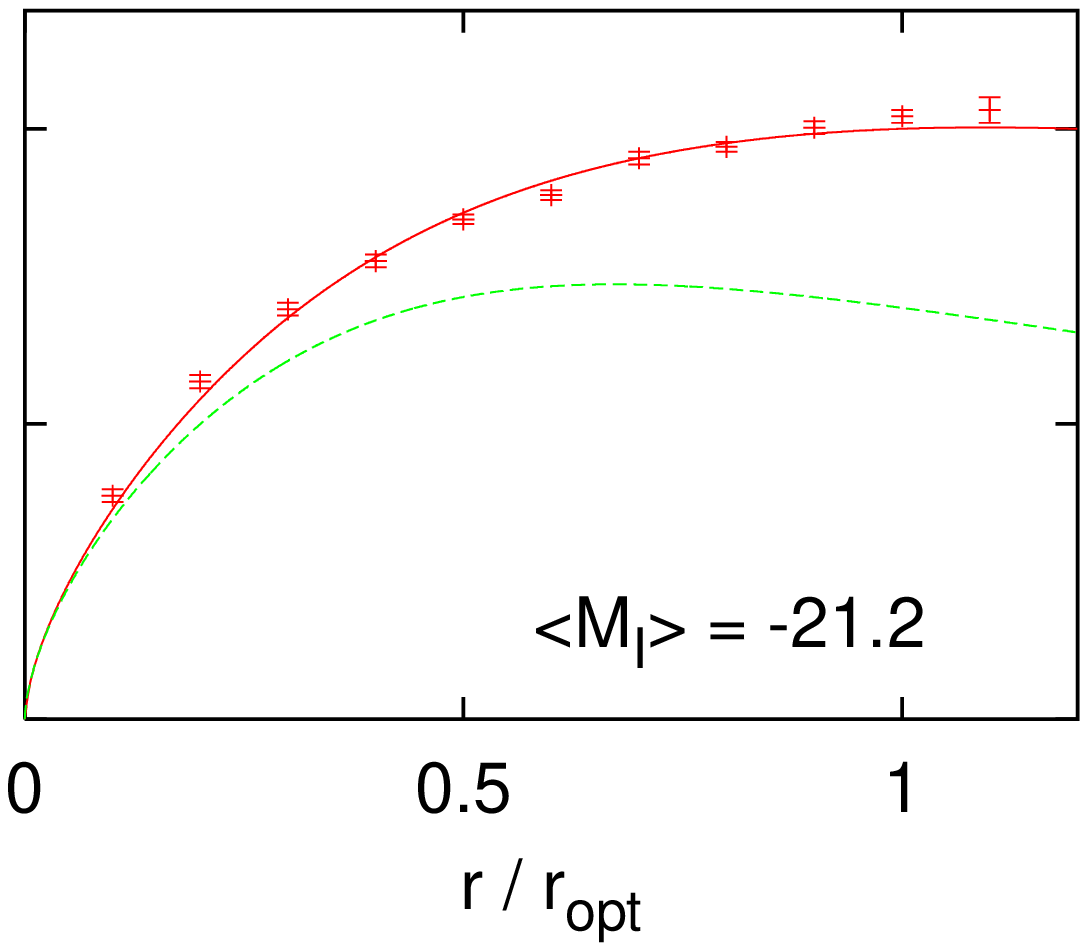}
\vspace{1.cm}
\\
\caption{ In each panel, the average value $\left< r_{\rm opt} {\cal
V}_{\rm opt} \right>$ over the corresponding spiral galaxies has been
used to get the scalar field mass $m \, = \, \alpha \sqrt{\gamma} /
(r_{\rm opt} \, {\cal V}_{\rm opt})$.  The best fit configurations
with a common value of the mass $m = 6 \times 10^{-24} \, {\rm eV}$
are shown in solid lines, together with the observations. Again, the
dashed line shows the contribution from the disk only.}
\label{fig:fig5}
\end{figure}
%
\vskip 0.1cm
For each individual galaxy, one can restore the physical value
of the rotation velocity by adjusting the free scaling parameter
$\bar{\sigma}(0)$. The mass of the scalar field is given by
\begin{equation}
m^{2} \, = \, {\displaystyle \frac{\alpha^{2} \, \gamma}
{r_{\rm opt}^{2} \; \beta \; {\cal V}^{2}(r_{\rm opt})}} \;\; .
\end{equation}
Here we defined the parameter
$\beta = {\cal V}_{\rm disk}^{2}(r_{\rm opt}) /
{\cal V}^{2}(r_{\rm opt})$, calculated by the code for each value of
the relevant parameters $\{ \alpha, \gamma \}$.  Of course, our
model can provide a powerful explanation for galactic rotation curves
only if all galaxies can be fitted simultaneously with the same value
of $m$, and therefore
approximately the same angular velocity $\omega$ (in principle one
could introduce a time-dependant effective mass, with slightly
different values at large and small redshifts; however, since galaxy
rotation curves have never been found to be redshift-dependant, we
discard this possibility and assume that $m$ is constant in space and
time, at least on observable scales).
In order to test this idea, it would be natural to use data
from individual galaxies; but in doing so, one would face back the
problems associated with large systematic uncertainties, which
motivate the PSS approach. In a first step, we plot $\chi^{2}$ as a
function of $m \; r_{\rm opt} \, {\cal V} (r_{\rm opt})$ in Fig.~\ref{fig:fig4}
and remark that the faint galaxies have a preferred mass while the
brighter ones in the two last panels provide only an upper bound.
Then, we assume that each synthetic universal rotation curve is
associated with a unique typical galaxy, with average optical radius
and velocity.  For each class of magnitude, we perform an average on
the sub--sample given by PSS -- see the tables in their Appendix D --
and find respectively
$\left< r_{\rm opt} {\cal V}(r_{\rm opt} \right> = 710$,
1100, 1900, 2500, 3200 and 4900~kpc~km~s$^{-1}$. Since we do not
employ the raw 600 galaxy data, we cannot give an error on these
numbers, nor can we make a precise prediction for the mass. However, plotting
now the $\chi^{2}$ as a function of $m$, we find that the first four
classes are perfectly compatible with a mass
$m \in [4 - 16] \times 10^{-24} \, {\rm eV}$ while the two others push
towards the lower--end of this interval with
$m \in [4 - 8] \times 10^{-24} \, {\rm eV}$.
\vskip 0.1cm
On Fig.~\ref{fig:fig5}, we plot the rotation curves obtained with
$m = 6 \times 10^{-24} \, {\rm eV} \simeq 5 \times 10^{-52}$ M$_{\rm P}$
and minimized over $\gamma$.  Although the $\chi^{2}$
values are approximately 1.5 bigger than those of the independent
best--fits of Fig.~\ref{fig:fig3}, the agreement with the data remains
quite good. The effect of fixing the mass is to obtain more radical
behaviors at $r_{\rm opt}$: for light galaxies, the rotation curves are
growing faster near the optical radius while for bright galaxies they are
even flatter. For all these models, the total mass is close to
$M = 5 \times 10^{10} \, {\rm M_{\odot}}$ whereas the ratio of the total
radius over the optical radius -- which respectively encompass 83\% of the
total and luminous mass -- varies between 4 for $<M_I> = - 18.5$ and 1.5
for $<M_I> = - 21.2$.


\section{Discussion and prospects.}
\label{sec:discussion}

In this work, we solved the Einstein and Klein--Gordon equations for a
free massive scalar field, in presence of a baryonic disk. Using the
universal curves of Persic, Salucci and Stel \cite{persic_salucci_stel}, 
which are based on
hundreds of galaxies, we showed that a galactic halo consisting in
such a Bose condensate could explain fairly well the rotation of
low--luminosity spiral galaxies. A single value of the mass, of order
$6 \times 10^{-24}$eV, is compatible with galaxies of different
magnitudes. The corresponding Compton wavelength $1/m \simeq
10^{-2}$ kpc is three orders of magnitudes smaller than the typical size
of spirals. Indeed, the spatial extension of a
self--gravitating field $\Phi$ is given approximately by
$1 / (\sqrt{\Phi(0)}~m)$, while the square root of the central field value
(expressed in Planck units) is comparable with the velocity of
orbiting particles (in units of $c$). Since we are dealing with speeds
${\cal V} \sim 100$~km~s$^{-1}~\sim 10^{-3} \, c$, there is really a
factor of $10^3$ between the Compton wavelength and the size of the halo.
Note that since $\omega \simeq m$, the light scalar field $\Phi$ rotates
in its internal space with a period of $\sim 30$ yr.

\vskip 0.1cm We conclude that scalar fields could be a nice
alternative to CDM halo models. Our positive results concerning the
rotation curves are strengthened by other astrophysical
considerations. First, a scalar field solves naturally the dynamical
friction issue for bared galaxies. Because it is completely smooth,
such an extended field cannot slow down the spinning bars observed at
the centers of many galaxies, as would a granular CDM medium. Second,
following Hu, Barkana and Gruzinov \cite{Hu}, an ultra--light scalar
field can avoid the excess of small--scale structure predicted by CDM
simulations near the galactic center.

\vskip 0.1cm Of course, several improvements are needed before
concluding that a scalar field is the best galactic dark matter
candidate on the market. First, it is necessary to extend the
comparison to various types of individual galaxy rotation curves, with
the drawback that more degrees of freedom must be included in
realistic modelizations of the baryonic components (gas, bulge,
...). This is however the only way to obtain better constraints on
$m$, and to find out whether a quartic coupling, not considered in
this analysis, improves the model. In some particular cases, it would
be worth taking into account excited field configurations, which seem
to be stable (due to charge conservation) and which predict
ultra--flat rotation curves far from the core, with small wiggles that
may have already been observed (see also the claim in \cite{Roscoe}
concerning possible existence of discrete dynamical classes for spiral
galaxy disks). Also, in order to get a better view of the rotation
curves in the vicinity of the core, especially for bright spiral
galaxies, further technical ingredients must be passed to the
equations, in order to distinguish the spherical symmetry of the halo
from the quasi two--dimensional distribution of the stars.

\vskip 0.1cm
Finally, it would be extremely interesting to plug such a complex
light scalar field into a general cosmological framework, and study
into details the growth of linear perturbations and the formation of
non--linear structures. The pioneering discussions on such cosmological
scenarios \cite{Hu,Kamion} are very promising and suggest that many
interesting developments on scalar field dark matter should arise in
the next years.

\section*{Acknowledgements}

We would like to thank R.~Taillet and J.-P.~Uzan for useful discussions.



\end{document}